\documentstyle[twocolumn,12pt,psfig]{article}
\begin{document}

\renewcommand{\baselinestretch}{1.5}
% List of macros for C.C.: the euc. isometries
\def\B{\cal{B}}
\def\L{{\cal L}}
\def\X{{\cal X}}
\def\Bg{{\cal B}_g}
\def\Cr{{\cal C}_r}
\def\SgBr{S_g^{\cal B}(r)}
\def\QgBr{Q_g^{\cal B}(r)}
\def\SgBx{S_g^{\cal B}(x)}
\def\BUBg{{\cal B}\cap {\cal B}_g}
\def\BUCr{{\cal B}\cap {\cal C}_r}
\def\BgUCr{{\cal B}_g\cap {\cal C}_r}
\def\BUBgUCr{{\cal B}\cap {\cal B}_g\cap {\cal C}_r}
\def\PgBl{{\cal P}_g^{\cal B}(l)}
\def\BUBgUXx{{\cal B}\cap {\cal B}_g\cap {\cal X}_x}
\def\VgB{V_g^{\cal B}}
\def\bea{\begin{eqnarray}}
\def\eea{\end{eqnarray}}
\title{\Large \bf Cosmic crystallography: the euclidean isometries
\\ } 
\author{A. Bernui~$^\ast$ %\\ 
\vspace{-0.5cm} \\
{\small Facultad de Ciencias, Universidad Nacional
de Ingenier\'{\i}a} \vspace{-0.5cm}  \\
{\small Apartado 139, Lima 31, Peru} \\ 
A.F.F. Teixeira~\thanks{bernui@fc-uni.edu.pe, teixeira@cbpf.br} %\\
\vspace{-0.5cm} \\
{\small Centro Brasileiro de Pesquisas F\'{\i}sicas} %\\
\vspace{-0.5cm} \\
{\small 22290-180 Rio de Janeiro -- RJ, Brasil} \\ \\
       }
\date{}
% \\ {\bf d:/meutex/cceuc.tex}

%
\maketitle
\begin{abstract} %#################################################
Exact \ expressions \ for \ probability densities
of \ conjugate \ pair \ separation \ in \ euclidean \
isometries are obtained, \ for the cosmic
crys- tallography.~These~are~the~theoretical~counter-
parts \ of \ the \ mean histograms~arising \ from
computer simulation of the isometries. \\
For completeness, \, also the isometries with fixed points
are examined, \, as well as the orientation reversing
isometries.
\end{abstract}
{\raggedright
\section{Introduction} }  \label{Intro} %##########################
\setcounter{equation}{0}
Various methods have been proposed to investigate the shape of the
universe, and cosmic crystallography (CC) is one of
them \cite{lelalu}.
As CC ponders, if the universe is multiply connected (MC) then
multiple images of a same cosmic object may be seen in the sky. 
The separations between these images are correlated by the geometry 
and the topology of the spacetime; so if one selects a catalog of
$n$ observed images of various cosmic objects and performs a
histogram of the $n(n-1)/2$ separations between these $n$ images, 
then the existing correlations must somehow show up. 

It was recently examined \cite{spikesI} in what respects the
histogram for a multiply connected observed universe should differ
{}from that of a simply connected (SC) one, with same geometry and
radius.
It was found that each isometry of the MC universe individually
imprints either a small localized deformity on the histogram of the
SC universe \cite{spikesI}, or a sharp spike if the isometry is a
Clifford translation \cite{spikesII}.

Since each observed universe model has a unique
pair \, separation \, normalized \, histogram, a \,
strategy to unveil the cosmic topology seems straightforward:
one should compare the histogram obtained from observational
astronomy with histograms obtained from computer simulated universe
models with prescribed topologies. 

Both types of histograms (observational and simulated) are infected
with statistical noises, and methods to reduce these noises are
desirable.
A suggestion was made, to replace the histogram related to the SC
component of the simulated model by an {\it exact} continuous
probability density function. 
For each geometry with constant curvature the corresponding 
function was then derived \cite{cc3mpf};  
however, appropriate functions were still lacking, to replace  
the histograms related to each isometry component \cite{tsct}. 

In the present study we derive some of these functions, namely the
normalized pair separation probability densities for the
{\it euclidean} \ isometries. 
Following the prescriptions of ref. \cite{spikesI}, if we now  
merge these new functions with the functions already given
in~\cite{cc3mpf} then we obtain noiseless normalized probability
densities more suitable for comparing with the normalized
observational histograms.
For completeness we also examine the euclidean isometries with
fixed point, as well as the orientation reversing isometries.

\section{$\!\!\!\!\!\!$ The euclidean isometries} \label{sec2} %####
\setcounter{equation}{0}

The isometries of the euclidean space $E^3$ are the (pure)
translations $t$, (pure) rotations $\omega$, screw motions $t\omega$,
(pure) reflections $\epsilon$, and glide reflections $\epsilon t$. 
While $t$, $\omega$, and $t \omega$ are orientation preserving,  
$\epsilon$ and $\epsilon t$ reverse the orientation of $E^3$; 
and while $t$, $t \omega$, and $\epsilon t$ act freely on $E^3$, 
$\omega$ and $\epsilon$ have fixed points. 
We study all five isometries for the sake of completeness, 
although cosmic crystallography is presently interested on the
orientation preserving isometries without fixed points, namely $t$
and $t \omega$ only.

Our general approach to investigate an isometry $g$ in cosmic
crystallography is first consider a solid ball $\B$ with radius $a$,
then apply $g$ to $\B$ thus producing a new solid ball $\Bg$, 
next consider the set of points $P$ $\in$ $\B$ whose corresponding 
transformed points $P_g$ are also in $\B$,  
finally randomly select a pair ($P$, $P_g$) and ask for the
probability $\PgBl dl$ that their separation lies between the values
$l$ and $l+dl$.
The normalization condition 
\bea                                                   \label{2.1}
\int_0^{\infty}\PgBl dl = 1
\eea
must be obeyed.

It is clear that the balls $\B$ and $\Bg$ need intersect, so the
separation $m$ between their centers $C$ and $C_g$ has to satisfy
\bea                                                  \label{2.2}
m < 2a \, .
\eea
The intersection $\BUBg$ is a rotationally symmetric solid lens 
whose diameter, thickness, and volume are (see figure 1) 
\bea                                                 \label{2.3}
D &=& \sqrt{4a^2-m^2}, \nonumber \\ 
T &=& 2a-m,  \\
V_{g}^{{\cal B}} &=& \frac{\pi}{12}(2a-m)^2(4a+m) \, \nonumber.
\eea
%\begin{figure} 
\hspace*{1.0cm}
\psfig{file=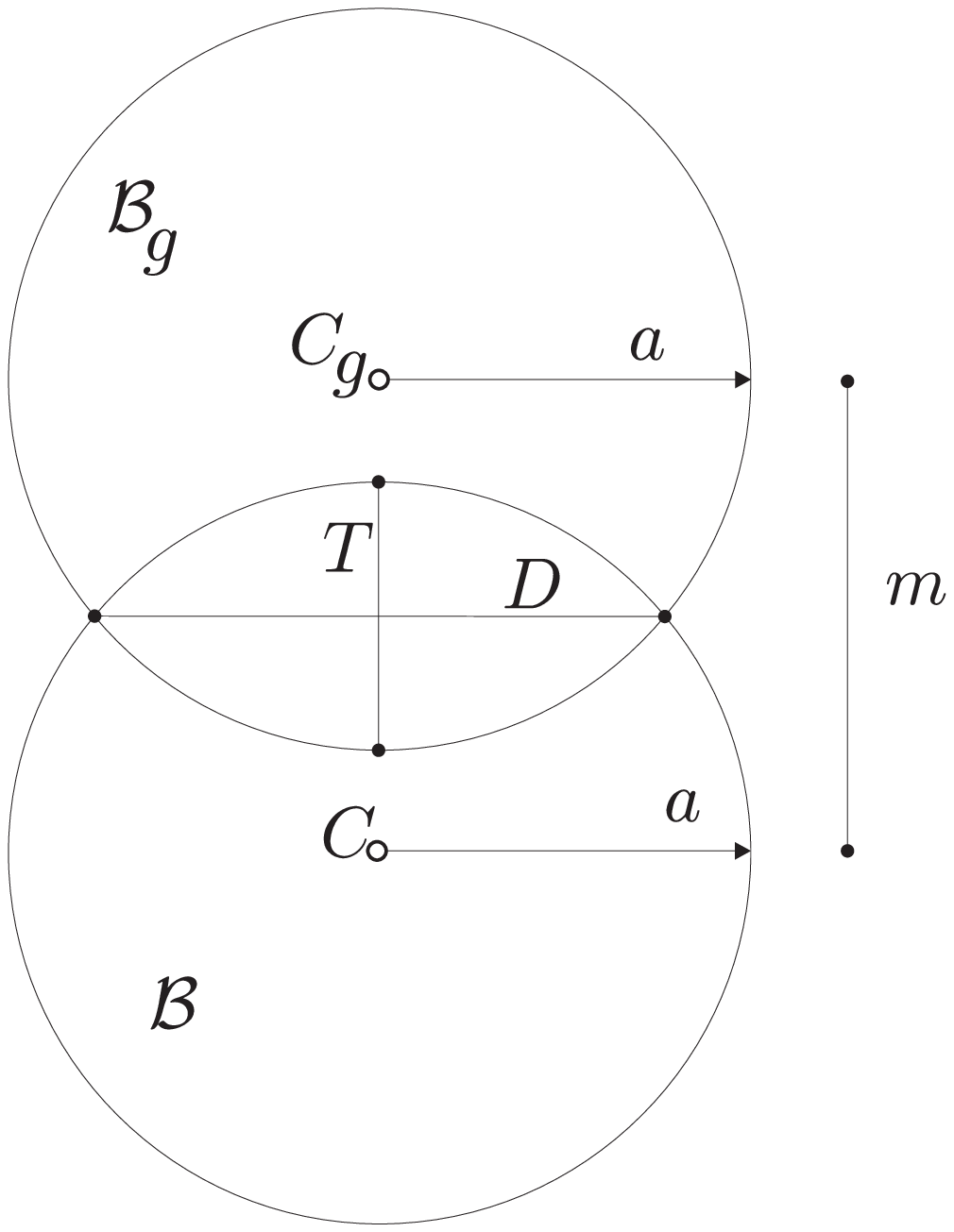,height=8cm}     %Figura0101010101010101010101

\vspace*{0.4cm}
\noindent
{\bf Figure 1} The solid balls $\B$ and $\Bg$ both with radius
$a$ intersect in a solid lens with diameter $D$ and
thickness $T$. $\hfill {\Box\,}$
%\caption{?}
%\end{figure}
%
\vspace*{-1cm}
%###################################################################
\section{$\!\!\!\!$ Translations, screw mo-tions,
and rotations} \label{sec3}
\setcounter{equation}{0} 
%
%###################################################################
{\bf Translations}

\noindent
If the euclidean space is subjected to a nonzero translation $t$, 
the probability density that a point $P$ be displaced a distance
$l$ is clearly
\bea                                                 \label{3.1}
{\cal P}_{t} (l)=\delta (l-t) \, , 
\eea
the Dirac delta; the density satisfies the normalization condition
(\ref{2.1}), and does not depend on the ball $\B$.
%
%###################################################################

\vspace*{0.3cm}
\noindent
{\bf Basics on screw motions of a solid ball}

\noindent
In $E^3$, imagine a straight line $\L$ 
placed at a distance $b$ from the center $C$ of the solid ball $\B$ 
with radius $a$; $\L$ and $\B$ may intersect ($b<a$), be tangent
($b=a$) or be disjoint ($b>a$); see figure 2.
\vspace*{0.4cm}
\hspace*{1cm}
\psfig{file=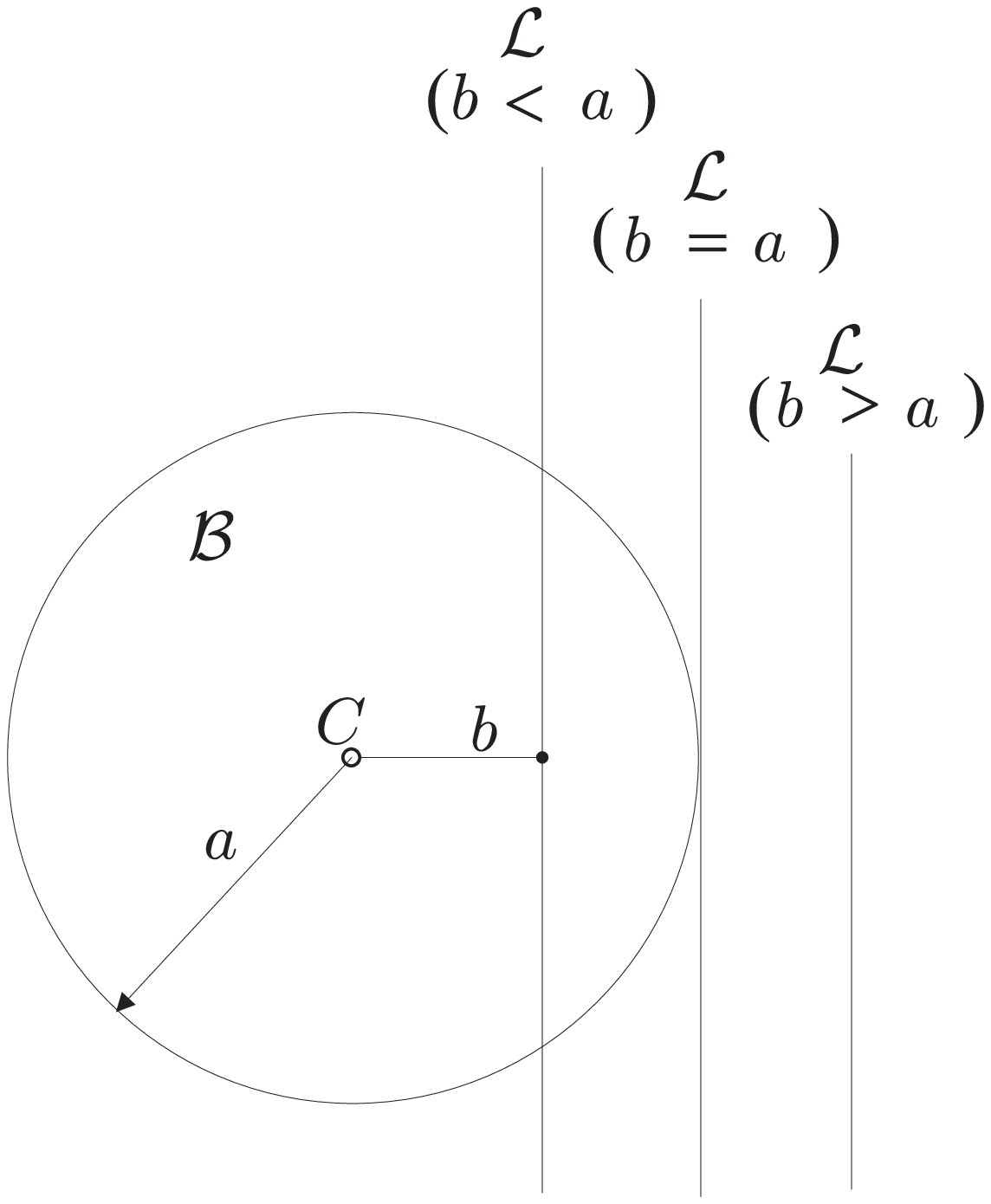,height=8cm}  %Figura 020202020202020202

\noindent
{\bf Figure 2} Relative positions of a solid ball $\B$ with radius
$a$ and a straight line $\L$ at a distance $b$ from the center
of $\B$. $\hfill {\Box\,}$

Now consider a screw motion $g$ of the ball, with nonzero
translation $t$ parallel to the line $\L$ and nonzero rotation
$\omega$ around the line;
for our purposes the senses of $t$ and $\omega$ are irrelevant, 
so for definiteness and simplicity we assume $t>0$ and
$0<\omega \leq \pi$.

The separation $m$ between the centers $C$ and $C_g$ of the
intersecting balls $\B$ and $\Bg$ is 
\bea                                                 \label{3.2}
m=\sqrt{t^2 + 4\,b^2{\sin}^2 \omega/2},
\eea
and the condition $m<2a$ implies the constraint 
\bea                                                 \label{3.3}
t^2 + 4\,b^2{\sin}^2 \omega/2<4a^2 
\eea
between the four independent parameters $a$, $b$, $t$, and $\omega$; 
see figure 3. 
\vspace*{0.4cm}
\hspace*{1cm}
\psfig{file=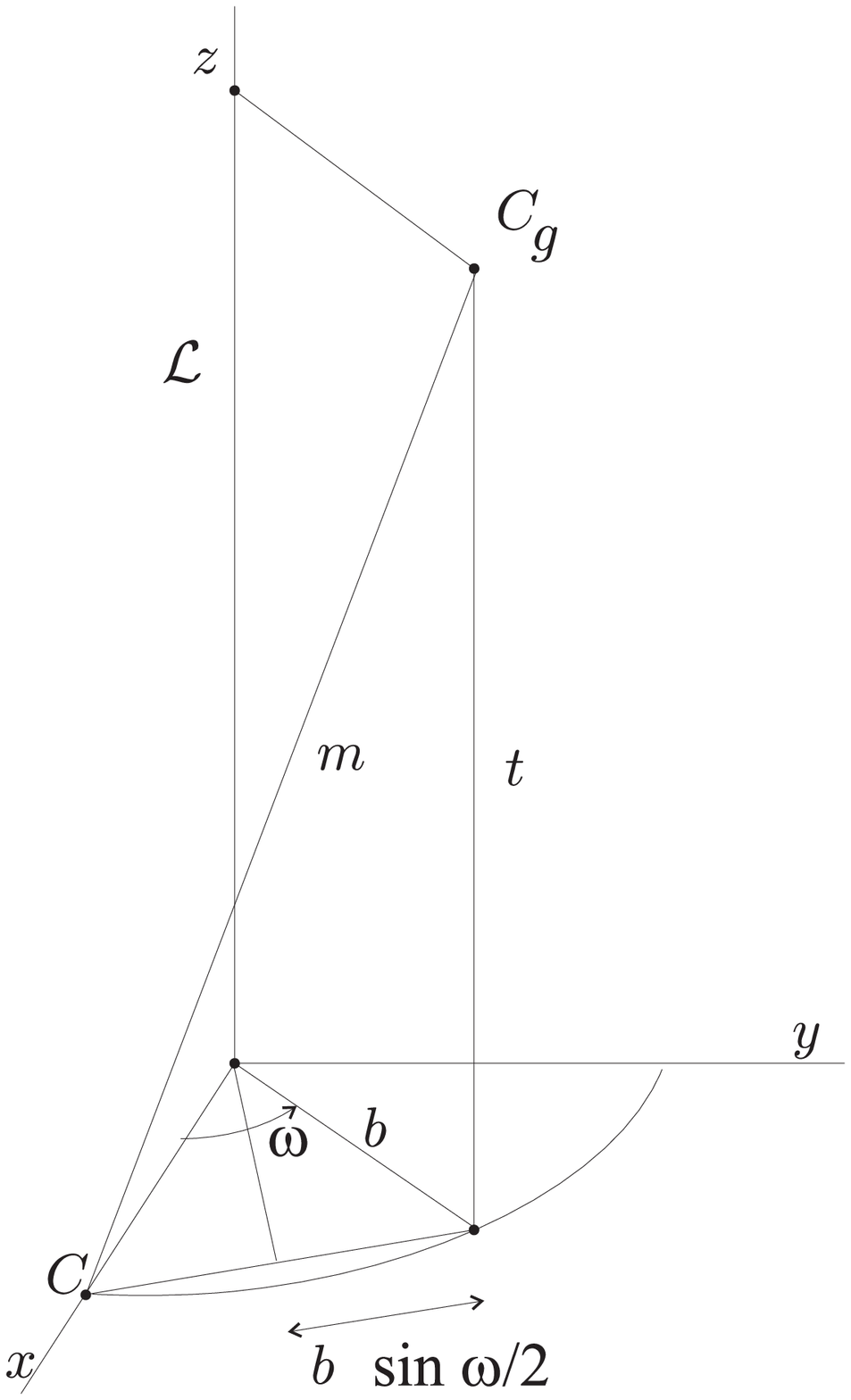,height=10cm}   %Figura0303030303030303030303

\vspace*{0.4cm}
\noindent
{\bf Figure 3} A solid ball (not displayed) with center $C$
rotates $\omega$ around the line $\L$ (the $z$ axis) and translates
$t$ parallel to the axis, eventually reaching the position  centered
at $C_g$\,; both centers $C$ and $C_g$ are at a distance $b$ from
the axis, and their separation $m$ must be smaller than the sum $2a$
of the two radii in order that the balls
intersect. $\hfill {\Box\,}$

\vspace*{0.3cm}
We next introduce the auxiliary variable $r$ and a function
$\QgBr$ which help simplify our study.  
Call $r$ the separation between the point $P$ (also $P_g$) and
the axis $\L$ (see figure 4);
it is related to  $l$ (the separation from $P$ to $P_g$) through
\bea                                                 \label{3.4}
l^2=t^2 + 4\,r^2{\sin}^2 \omega/2 \, \, ; 
\eea
for each isometry ($t,\omega$) this is a bijective relation between 
$l$ and $r$, so the probability density $\PgBl$ 
of finding a $g$-pair with mutual separation $l$ parallels 
the akin probability density $\QgBr$ of finding a $g$-pair 
whose members (both in $\B$) are at a distance $r$ from the axis of
the motion: 
\bea                                               \label{3.5} 
\PgBl dl=\QgBr dr.
\eea
\hspace*{0.5cm}
\psfig{file=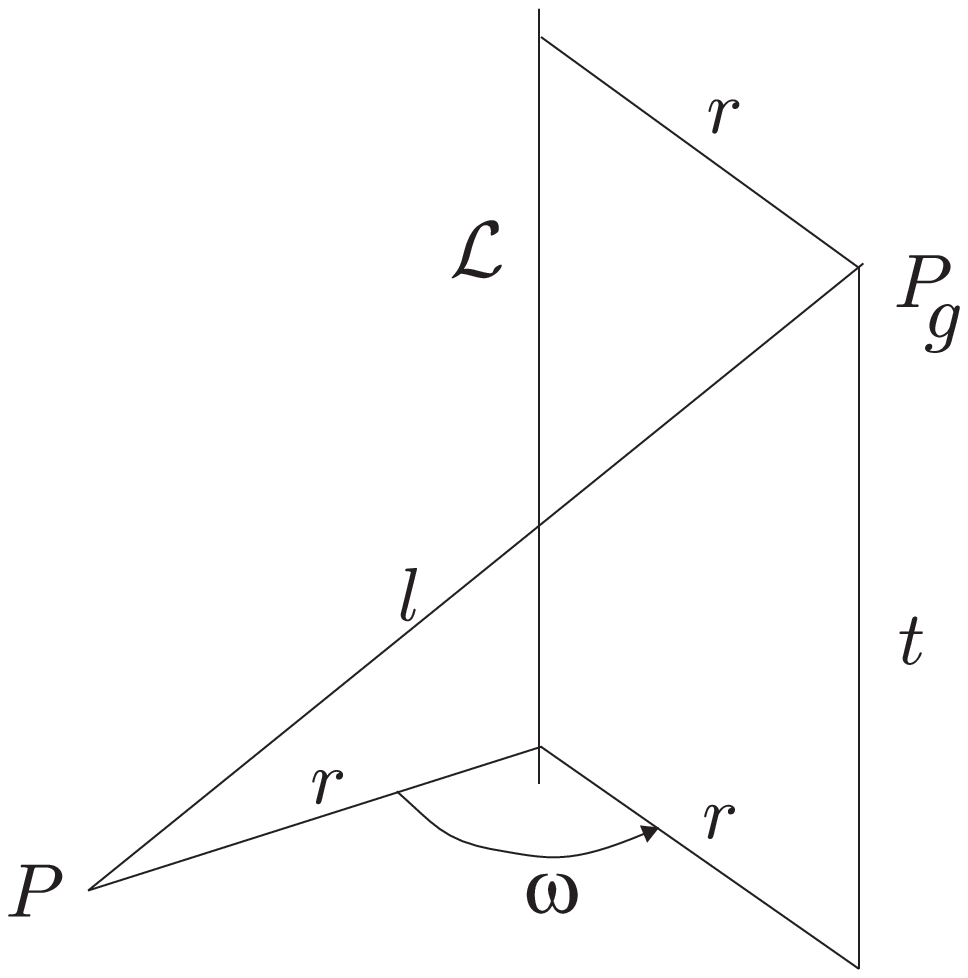,height=7cm,width=7cm}   %Figura040404040404

\noindent
{\bf Figure 4} Under the screw motion $g$ with axis $\L$\,,
translation $t$\,, and rotation $\omega$\,, a point $P$ separated
$r$ from $\L$ moves to the position $P_g$ at the distance $l$
given by eq. $(\ref{3.4})$. $\hfill {\Box\,}$

\vspace*{0.3cm}
Clearly the normalization condition
\bea                                               \label{3.6}
\int_{0}^{\infty}\QgBr dr = 1 
\eea
is also to be satisfied. Once we obtain 
$\QgBr$, and since from (\ref{3.4}) we have 
\bea                                                \label{3.7}
r = \frac {\sqrt{l^2 - t^2}} { 2\,{\sin}\,{\omega/2} }\, \,  ,
\eea
then we will finally compute 
\bea                                                \label{3.8}
\PgBl= \frac{dr}{dl}\,{\cal Q}_{g}^{\cal B}[r(l)].
\eea

To obtain $\QgBr$ we first consider a sufficiently long 
cylinder $\Cr$ with radius $r$ and axis $\L$. 
A short reflection gives that the probability density $\QgBr$ 
is linearly proportional to the area $\SgBr$ of the intersection 
$\BUBgUCr$; 
the coefficient of proportionality is the inverse of the volume 
$\VgB$ of the intersection $\BUBg$\,, eq.(\ref{2.3}): 
\bea                                                \label{3.9} 
\QgBr=
\frac{area (\BUBgUCr)}{vol (\BUBg)}=
\frac{\SgBr}{\VgB}. 
\eea 
To find the area $\SgBr$ we first assume the line $\L$ 
(the axis of the  screw motion) along the cartesian $z$-axis; the
cylinder $\Cr$ with axis $\L$ and radius $a$ then has equation 
\bea                                                \label{3.10} 
x^2 + y^2 = r^2. 
\eea 
Still without loss of generality we take the center $C$ of the solid
ball $\B$ with radius $a$ at the cartesian position ($b$, 0, 0); 
the points of $\B$ then satisfy 
\bea                                               \label{3.11} 
(x-b)^2 + y^2 + z^2 \leq a^2. 
\eea 
Finally the center $C_g$ of the solid ball $\Bg$ is at the cartesian 
position ($b$ cos$\,\omega$, $b$ sin$\,\omega$, $t$), so the points
of $\Bg$ satisfy
\bea                                               \label{3.12}
\hspace{-0.5cm}
(x - b\cos\omega)^2 \!&+& \!(y - b\sin\omega)^2 \nonumber \\
                      &+& (z-t)^2 \leq a^2 \, .
\eea
$\SgBr$ is then the area of the surface whose points ($x,y,z$) 
satisfy (\ref{3.10}), (\ref{3.11}), and (\ref{3.12}) simultaneously. 
To visualize this surface $\BUBgUCr$ 
we first consider the surface $\BUCr$ , then the similar 
surface $\BgUCr$, and finally the {\sl combined} 
intersection $(\BUCr)\cap(\BgUCr)$. 

The form of the intersection $\BUCr$ depends on the relative 
values of $a, b$, and $r$ (see figure 5):

\noindent
(i) if $0<r<a-b$ then all generatrices of $\Cr$ penetrate into $\B$:
the intersection is a topological ring (an annulus, a compact
cylinder, a disc with a hole in it);

\noindent
(ii) if $|a-b|<r<a+b$ then only a part of the generatrices of $\Cr$
goes through $\B$: the intersection is a topological disc;

\noindent
(iii) if $a+b<r$, or if $0<r<b-a$, then there is no intersection. 

\vspace*{0.4cm}
\psfig{file=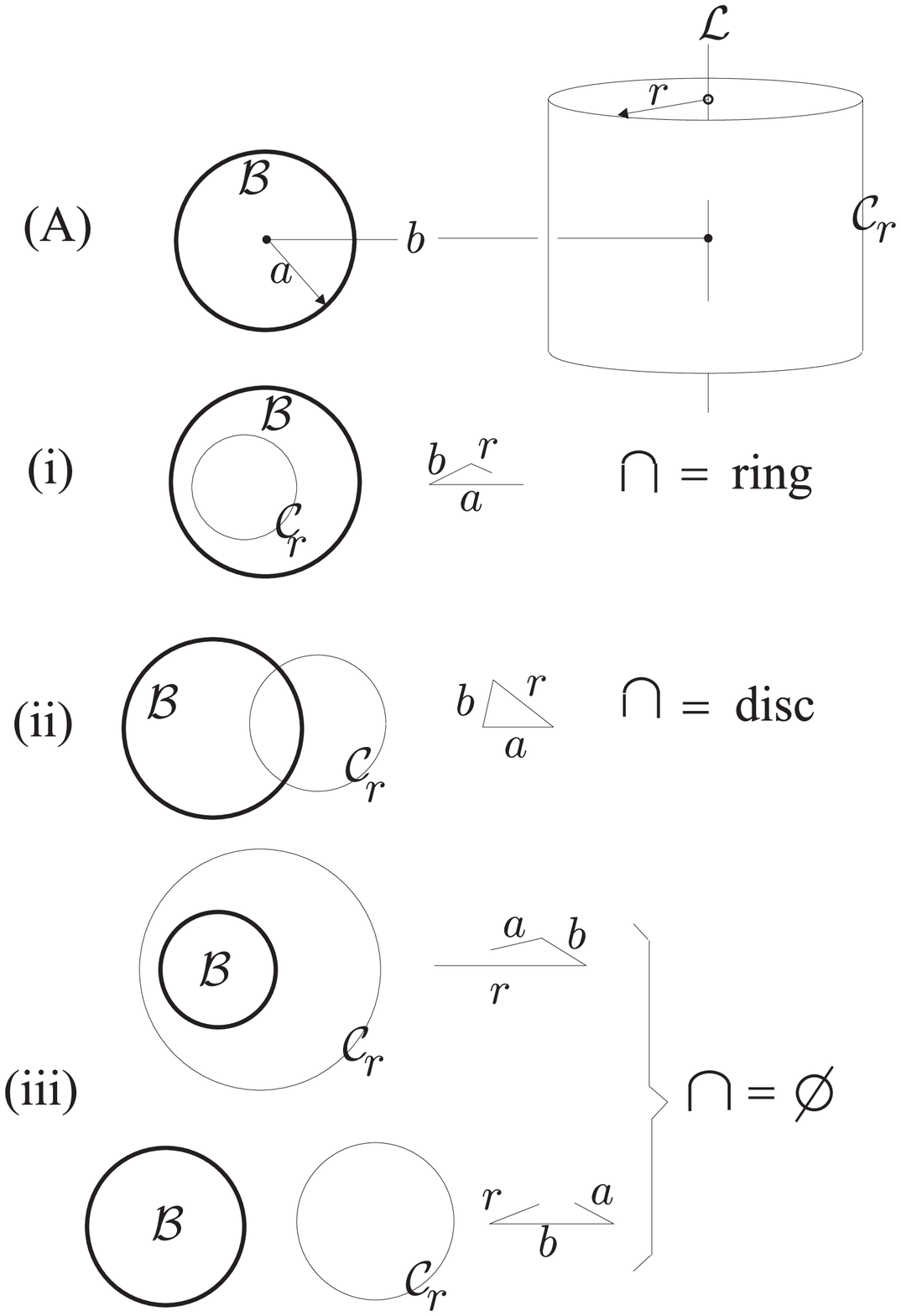,height=11cm}   %Figura05050505050505050505

\vspace*{0.4cm}
\noindent {\bf Figure 5}
{\bf (A)} The solid sphere $\B$ and the sufficiently long
cylindrical surface $\Cr$\,, in perspective; in the other drawings 
the line of sight is the vertical;

\noindent
{\bf (i)} the intersection $\BUCr$ is a topological ring when
$0<r<a-b$\,
(loosely saying, $a$ is too large);

\noindent
{\bf (ii)} $\B$ and $\Cr$ intersect in a topological disc when
$|a-b|<r<a+b$ (then $a, b$, and $r$ may form a triangle);

\noindent
{\bf (iii)} $\B$ and $\Cr$ do not intersect when $a+b<r$ nor
when $0<r<b-a$ (loosely saying, either $r$ or $b$ is too
large). $\hfill {\Box\,}$

\vspace*{0.3cm}
Since the intersections $\BUCr$ are drawn on the cylinder
$\Cr$ itself, we use the cylindrical coordinates $\phi$ (the
azimuthal angle) and $z$ (altitude) to visualize them. A ring-like
intersection $\BUCr$ is then bounded by the two curves (see figure 6) 
\bea                                               \label{3.13}
z=\pm \sqrt{a^2-b^2-r^2+2br\cos{\phi} } \, , 
\eea 
where $\phi \in (-\pi,\pi]$; the positive curve oscillates between
the extreme values
\bea                                               \label{3.14}
z_{max} &=& \sqrt{a^2 - (b-r)^2} \, , \nonumber \\
z_{min} &=& \sqrt{a^2 - (b+r)^2} \, . 
\eea 
\hspace*{0.3cm}
\psfig{file=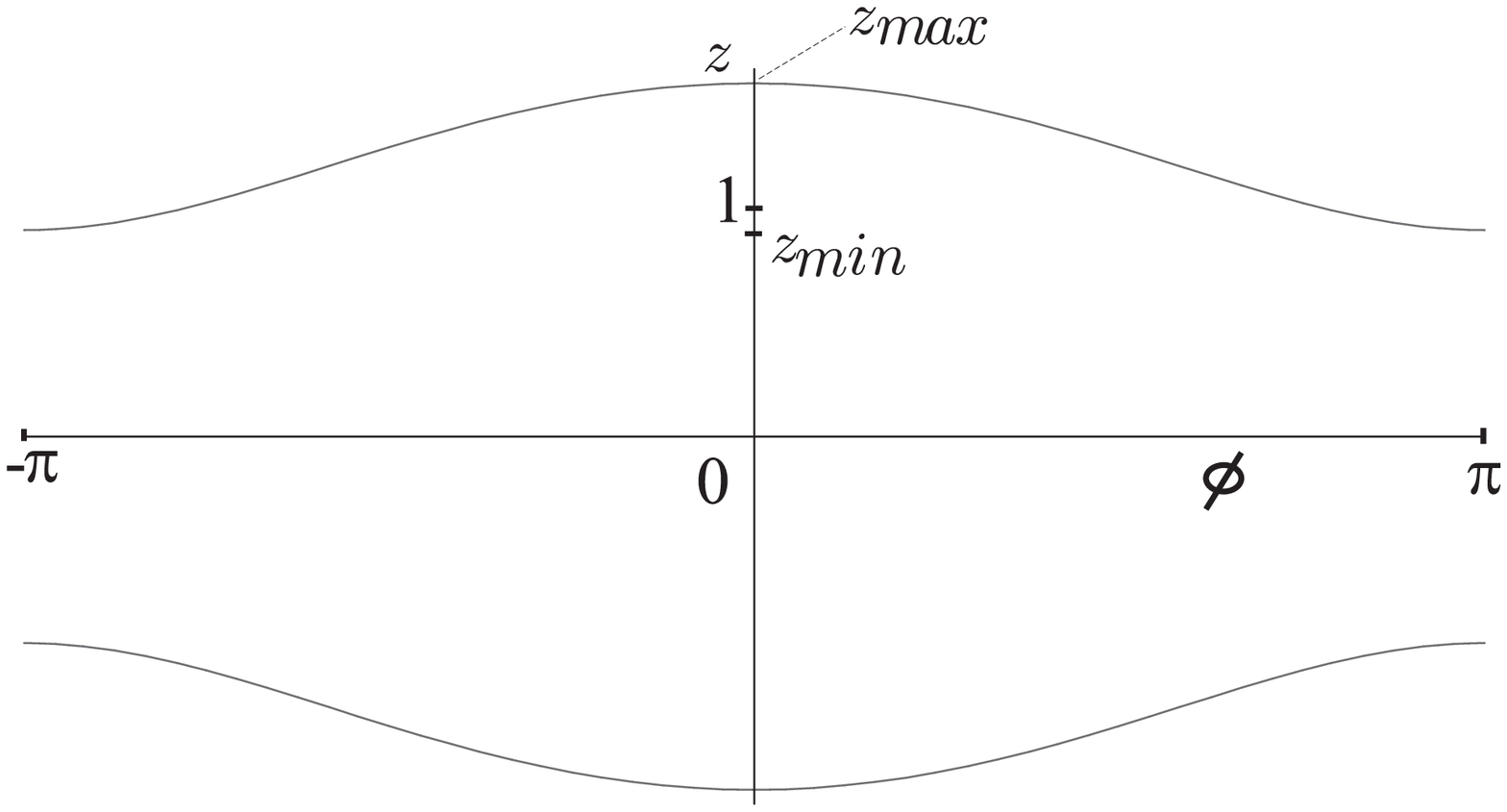,width=7cm}   %Figura06060606060606

\noindent {\bf Figure 6} The ring-like intersection $\BUCr$\,,
eq.(\ref{3.13}), when $a=4$\,, $b=19/5$\,, and $r=1/10$\,. 
The boundaries at $\phi=-\pi$ and $\phi=\pi$ are identified. 
The upper and lower boundaries are not
sinusoidal. $\hfill {\Box\,}$

\vspace*{0.3cm}
In the disc-like intersection $\BUCr$ (see figure 7) the surface
is again read from (\ref{3.13}), but now $\phi
\in [-{\phi}_{max},{\phi}_{max}]$
with 
\bea                                                \label{3.15}
{\phi}_{max}={\cos}^{-1}\,\,\,\frac{b^2+r^2-a^2}{2br};
\eea 
while $z_{max}$ is still given in (\ref{3.14}), $z_{min}=0$ now. 

\vspace*{0.4cm}
\hspace*{0.3cm}
\psfig{file=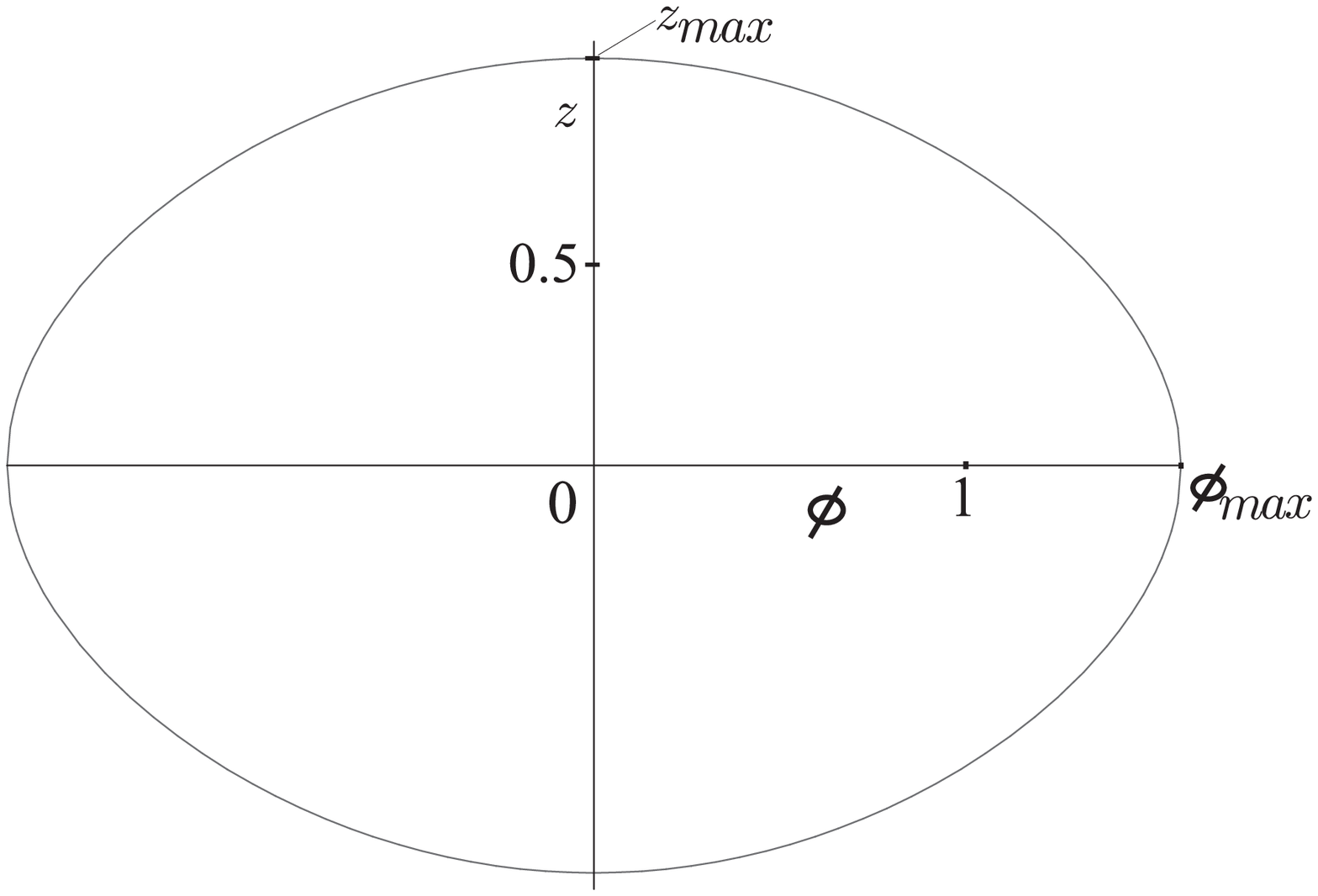,width=7cm}   %Figura07070707070707

\noindent {\bf Figure 7} The disc-like intersection $\BUCr$\,,
eq.(\ref{3.13}), when $a=1$ and $b=r=1/\sqrt{2}$\,; it is an oval 
(not an ellipse) centered at $\phi=z=0$\,, with $\phi_{max}=\pi/2$ 
and $z_{max}=1$\,. $\hfill {\Box\,}$

\vspace*{0.3cm}
A disc-like intersection $\BgUCr$ is a $g$-transported copy of
the corresponding intersection $\BUCr$, parallelly dragged on  
the cylinder: while $\BUCr$ is centered at ($\phi=0,z=0$), 
the center of $\BgUCr$ is at ($\phi=\omega,z=t$).  

Similar statements are true for the ring-like intersections 
$\BgUCr$. By inspection it is then easy to visualize the form of the 
{\sl combined} intersection $\BUBgUCr$ 
in both disc-like and ring-like cases. Since $\phi$ has a cyclic
character, the nonempty intersection of two discs on $\Cr$ can be
either 1 or 2 discs, while two rings intersect in either 1 or 2
discs, or 1 ring (see figures 8 and 9). 

\vspace*{0.4cm}
\hspace*{0.3cm}
\psfig{file=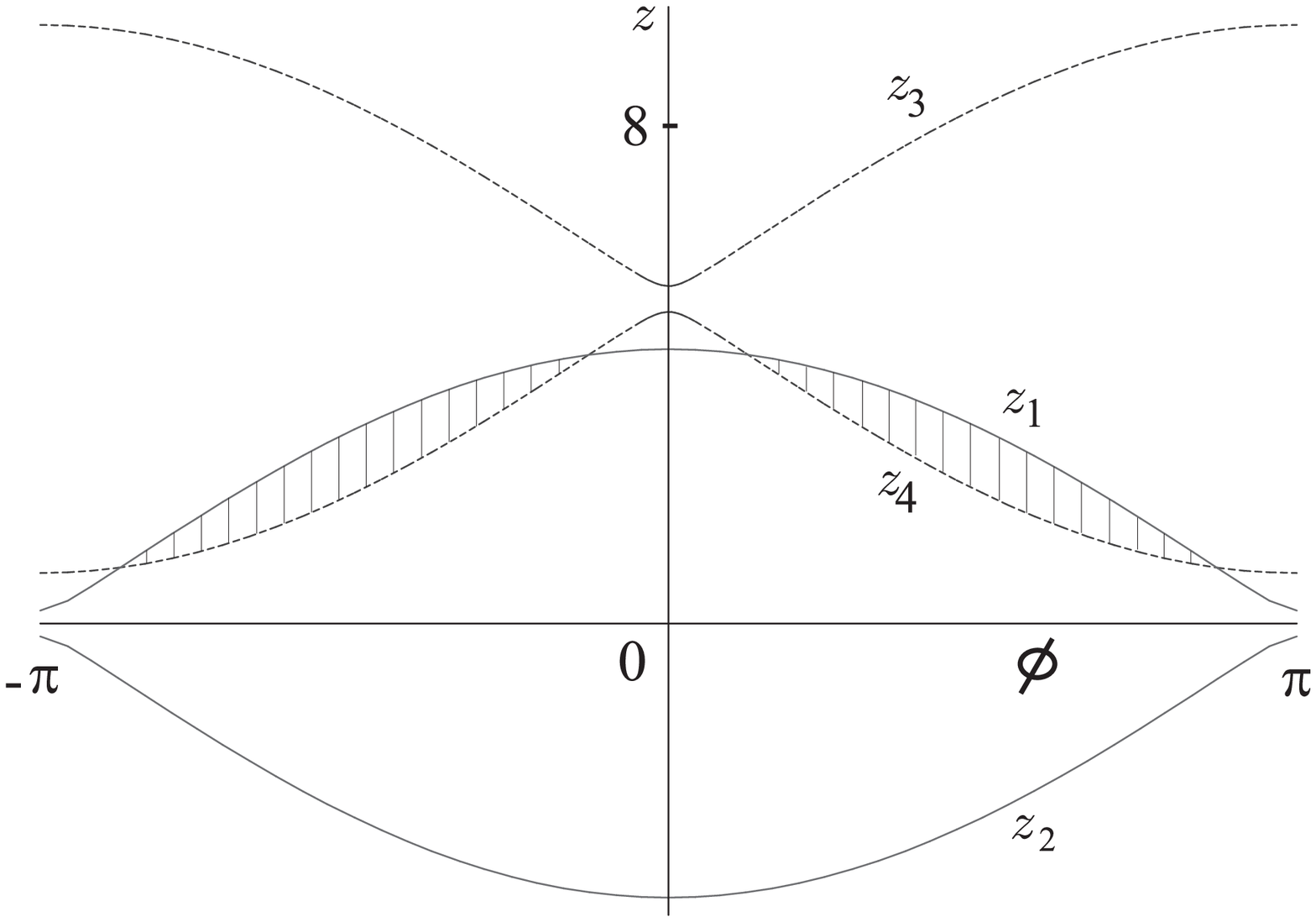,width=7cm}   %Figura0808080808

\noindent
{\bf Figure 8} Instance of disconnected intersection $\BUBgUCr$
(dashed areas) when $\BUCr$ (and $\BgUCr$) are ring-like; 
here $a=4.505\,, b=2\,, r=2.5\,, t=5.3$\,, and
$\omega=\pi$\,. $\hfill {\Box\,}$

\vspace*{0.6cm}
\hspace*{0.3cm}
\psfig{file=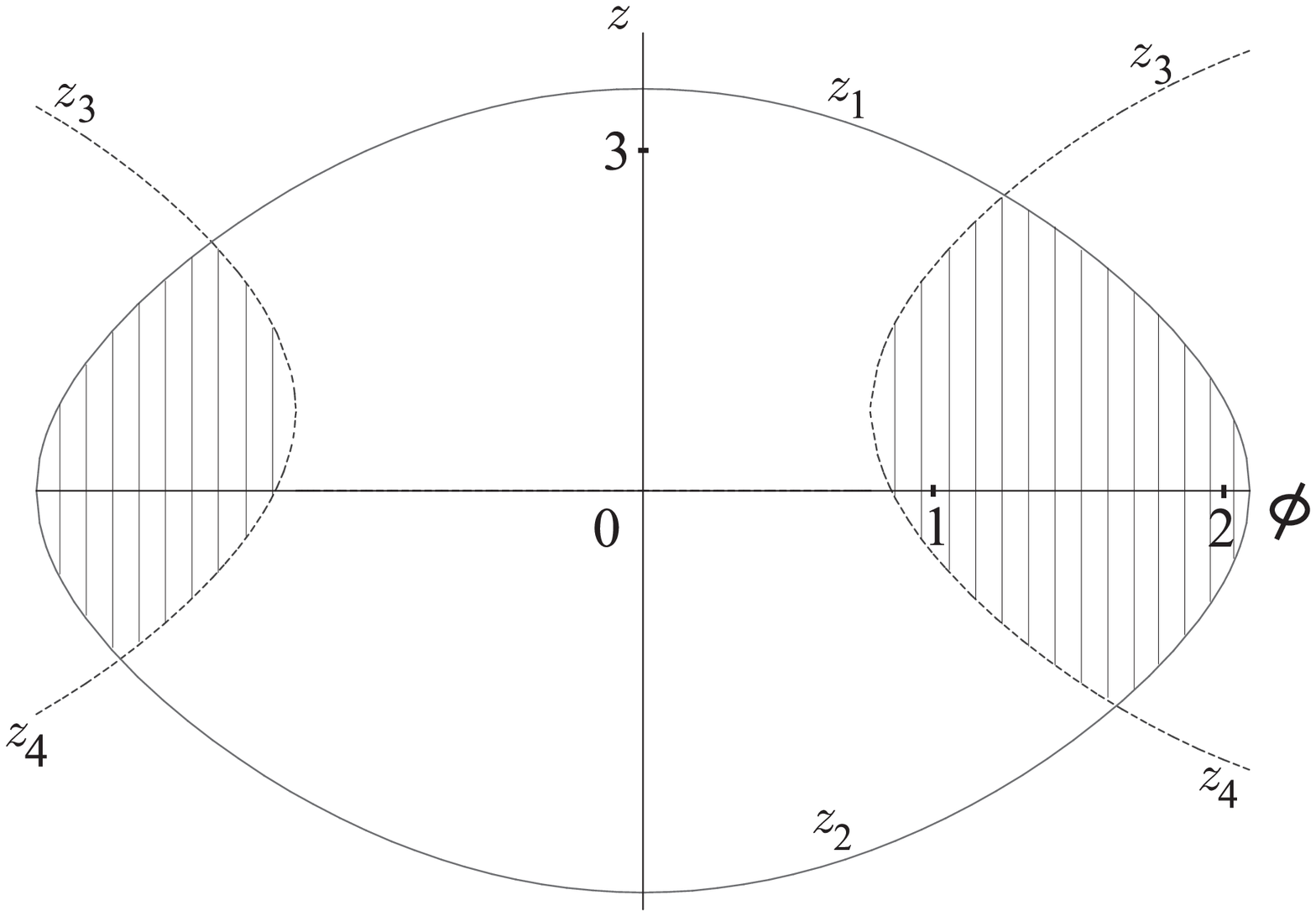,width=7cm}   %Figura0909090909090909

\noindent
{\bf Figure 9} Instance of disconnected intersection $\BUBgUCr$
(dashed areas) when $\BUCr$ (and $\BgUCr$) are disc-like; 
here $a=3.5\,, b=r=2\,, t=0.7$\,, and
$\omega=14\pi/15$\,. $\hfill {\Box\,}$

\vspace*{0.3cm}
We are now ready to evaluate the area  $\SgBr$ of the combined
intersection $\BUBgUCr$; given the fixed values 
of $a, b, t$, and $\omega$, then for each value of $r$ we need to
integrate the differential 
\bea                                                 \label{3.16}
d\SgBr={\Delta}z.r d\phi \, ,
\eea 
where ${\Delta}z(r,\phi)$ is the varying height of the intersection,
and where the limits of integration in $\phi$ may vary with $r$
(see eq.(\ref{3.15})).

As is evident from figures 8 and 9, the combined intersection is
always bounded by one of the two curves
\bea                                                 \label{3.17}
z_{1}(\phi)=\sqrt{a^2-b^2-r^2+2br\cos{\phi} } \,\, , 
\eea
\bea                                                 \label{3.18}
\!\!\!\!\!\!\!z_{3}(\phi) \!=\! t \!+\! \sqrt{a^2\!-\!b^2
\!-\! r^2 \!+\! 2br \cos(\!\phi \!-\! \omega\!) },
\eea
{}from the upper side, and one of the two curves 
\bea                                                 \label{3.19}
z_{2}(\phi)=-z_{1}(\phi)\,\, , 
\eea
\bea                                                 \label{3.20}
\!\!\!\!\!\!\!z_{4}(\phi) \!=\! t \!-\! \sqrt{a^2\!-\!b^2
\!-\! r^2 \!+\! 2br \cos(\!\phi \!-\! \omega\!) },
\eea 
{}from the lower side. Further, we only have nonempty combined
intersection when $z_{1}(\phi)>z_{4}(\phi)$. We then obtain 
\bea                                                 \label{3.21}
\hspace{-1.0cm} \SgBr \!\!\! &=&
r \int_{-\phi_{max}}^{\phi_{max}}\Theta(z_1-z_4) \nonumber \\
\hspace{-1.5cm} \times && \hspace{-1.0cm}
\Bigl[min(z_1,z_3)-max(z_2,z_4)\Bigr]d\phi  \, ,
\eea 
with $\phi_{max}$ given by (\ref{3.15}) for disc-like
intersections, and $\phi_{max}=\pi$ for ring-like intersections;
$\Theta$ is the step function with values 0 and 1. 
%
%###################################################################

\vspace*{0.3cm}
\noindent
{\bf Elliptic integrals }

\noindent
As is seen from eqs. (\ref{3.17})-(\ref{3.21}), 
the area $\SgBr$ involves terms of the form 
\bea                                              \label{3.22}
\!\!\!\!\! A(r) \!=\!\! \int_0^{\alpha (r)} \hspace{-0.5cm}
\sqrt{a^2 \!-\! b^2 \!-\! r^2 \!+\! 2br\cos\phi} \, d\phi ,
\eea
with $\cos(\phi-\omega)$ sometimes replacing $\cos\phi$ in the
integrand.
To evaluate these terms we define 
%???
%\begin{equation}
%\Theta(s) \equiv \left\{
%\begin{array}{ll}
%1 & \mbox{si $s > 0$} \\
%0 & \mbox{si $s < 0$}
%\end{array}
%\right .
%\end{equation}
%???
%
\begin{equation}                               \label{3.23}
\begin{array}{ccc}
f^{2}(r) &=& a^2-(b-r)^2 \, , \\
         & &                  \\
  k^2(r) &=&
\frac{\mbox{$4br$}}{\mbox{$f^{2}(r)$}} \, ,
\end{array}
\end{equation}
after noting that the condition $f^{2}(r)>0$ is satisfied by all
nonempty intersections $\BUCr$. 
Then introduce the (incomplete) elliptic integral of $2^{nd}$ kind 
\bea                                                  \label{3.24}
E(\gamma , k)=\int_0^{\gamma}\frac{\sqrt{1-k^2x^2}}
{\sqrt{1-x^2}}dx \, ,
\eea
and finally have the integral (\ref{3.22}) expressed as 
\bea                                                \label{3.25}
A(r)=2f(r)E\Bigl(\,\sin \frac{1}{2}\alpha(r),\, k(r)\,\Bigr) \; ,
\eea
where $f=\sqrt{f^2}$ and $k=\sqrt{k^2}$ \, . 
When $\gamma=1$ in (\ref{3.24}), or equivalently $\alpha=\pi$ in
(\ref{3.22}), we have the complete elliptic integral of $2^{nd}$ kind 
\bea                                                \label{3.26}
E(k)=E(1,k)\: ;
\eea
the functions $E(k)$ only appear in the ring-like intersections. 

Which of $z_1$ or $z_3$ is minimum in (\ref{3.21}), 
and which of $z_2$ or $z_4$ is maximum  
and whether $z_{1}>z_4 \; $or$\; z_{1}<z_4$\,, usually depends on
the angle $\phi$; clearly also the values of the parameters 
$a, b, t,$ and $\omega$\, are relevant, well as the radius $r$. 
%
%##################################################################

\vspace*{0.3cm}
\noindent
{\bf A sample screw motion with $b \neq 0$}

\noindent
We consider the case with
$a=t=1/\sqrt{2},\;  b=1/\sqrt{8},\; \omega=\pi$;
the intersection $\BUCr$ is then ring-like for 
$0<r<1/\sqrt{8}$  \, (equivalently $1/\sqrt{2}<l<1$ ), 
is disc-like for $1/\sqrt{8}<r<3/\sqrt{8}$  (or $1<l<\sqrt{5}$), 
and is empty for larger values of $r$ and $l$. 
The combined intersection $\BUBgUCr$ has a 
ring-like regime for $0<r<1/\sqrt{8}$ \,\,$(1/\sqrt{2}<l<1)$ \, ,
then a two separate discs regime for $1/\sqrt{8}<r<1/2$\, \,
%
%%%%%%%%%%%%%%%%%%%%%%%%%%%%%FIN%%%%%%%%%%%%%%%%%%%%%%%%%%%%%%%%%
%
$(1<l<\sqrt{3/2})$, and is empty for larger $r$ and $l$.

We first consider the ring-like regime: preliminarly find, from
(\ref{3.23}),
\begin{equation}                               \label{3.27}
\begin{array}{ccc}
f^{2}(r) &=&
\frac{\mbox{$ 3 $}}{\mbox{$ 8 $}} +
\frac{\mbox{$ r $}}{\mbox{$ \sqrt{2} $}} - r^2 \, , \\
         & &                  \\
k^{2}(r) &=&
\frac{\mbox{$ \sqrt{2} \, r$}}{\mbox{$ f^2 $}} \, ,
\end{array}
\end{equation}
and the area (\ref{3.21}) of the intersection
$\BUBgUCr$ ,
\bea                                              \label{3.28}
\SgBr=r\Bigl(8\,f\,E(k)-\pi\sqrt{2}\Bigr) \;\; ;
\eea
now since 
\bea                                               \label{3.29}
r=\frac{1}{2}\sqrt{l^2-1/2}\,, dr/dl=l/(4r)\,,
\eea
the probability density (\ref{3.8}) is, for $1/\sqrt{2}<l<1$ , 
\bea                                               \label{3.30}
\PgBl=\frac{l}{4\VgB}\Bigl(8\,f\,E(k)-\pi\sqrt{2}\,\Bigr) \; ,
\eea
where the volume of $\BUBg$ is 
\bea                                               \label{3.31}
\VgB=\pi (4\sqrt{2}-5)/12 \; .
\eea
Clearly $f(r)$ and $k(r)$ in (\ref{3.30}) need consider the
dependence $r(l)$ given in (\ref{3.29}).

To describe the two-discs regime we further define 
\bea                                               \label{3.32}
g(r) &=& \sqrt{1/(2r)^2-1} \, ,     \nonumber \\
g_{\pm}(r) &=& \sqrt{\frac{1 \pm g}{2}} \, , 
\eea
then obtain for the area (\ref{3.21}) 
\bea                                               \label{3.33}
\SgBr &=& 8\,r\,f\,\Bigl(E(g_{+},k)-E(g_{-},k)\Bigr) \nonumber \\
&+& \sqrt{2}r(2{\cos}^{-1}g-\pi) \; ,
\eea
and finally, for $1<l<\sqrt{3/2}$ , the probability density 
\bea                                               \label{3.34}
\PgBl &=& \frac{l}{4\VgB} \Bigl[ 8f\Bigl(E(g_{+},k)-E(g_{-},k)\Bigr)
\nonumber \\
&+& \sqrt{2}\,(2\,{\cos}^{-1}g-\pi) \Bigr] \; .
\eea
In (\ref{3.34}) we clearly need replace $r$ by its value $r(l)$ in  
(\ref{3.29}) in the functions $f, g, g_{+}, g_{-}$, and $k$.
In figure 10 we reproduce the function $\PgBl$ for this sample 
screw motion, for the whole interval $1/\sqrt{2}<l<\sqrt{3/2}$.  

\vspace*{0.5cm}
\hspace*{0.25cm}
\psfig{file=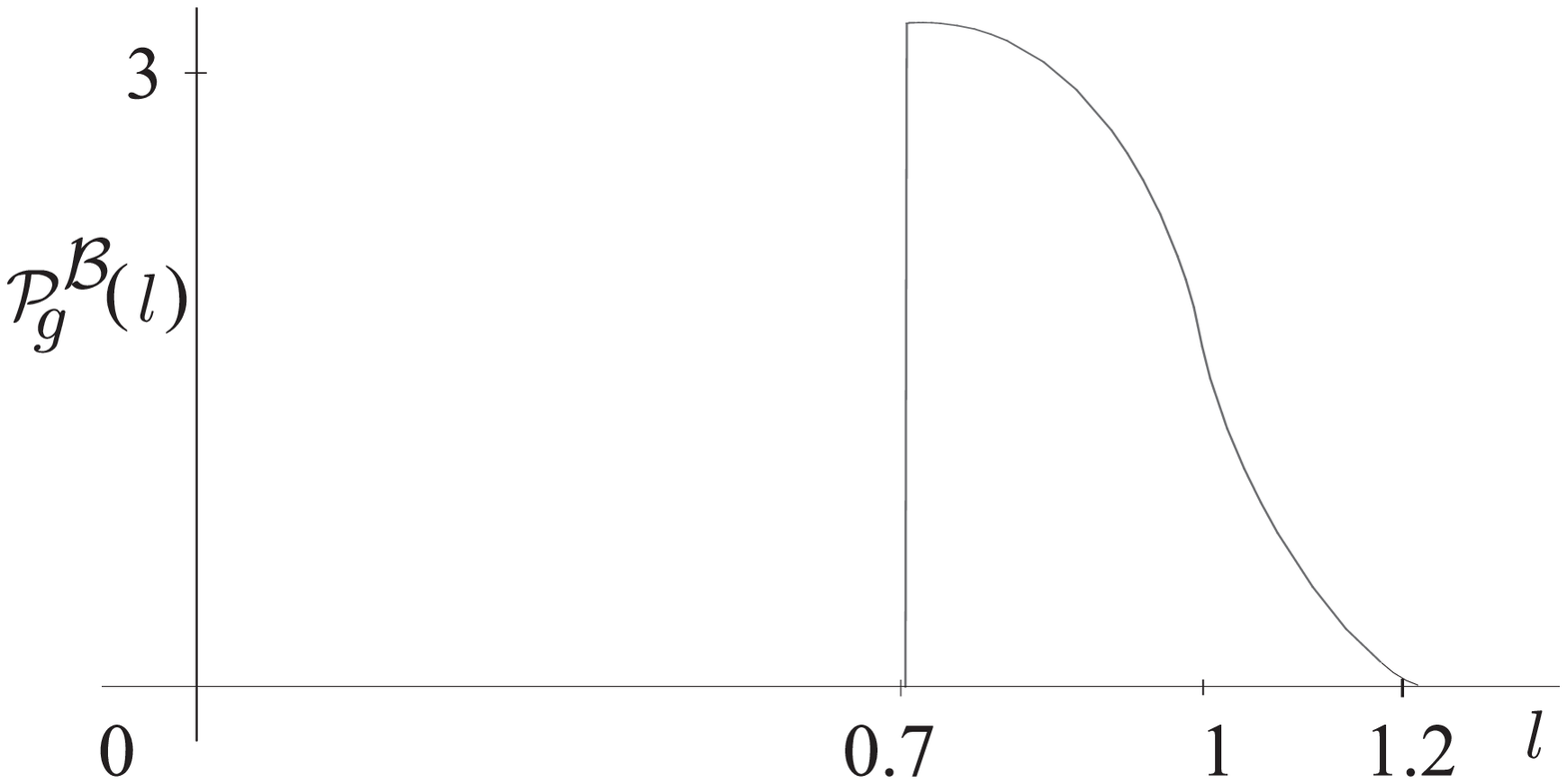,width=7cm}   %Figura10101010101010101010

\vspace*{0.4cm}
\noindent
{\bf Figure 10} The probability density $\PgBl$ for the screw motion
of the solid ball $\B$ with  $a=2b=t=1/\sqrt{2}$\,, and $\omega=\pi$.  
For $l \in [1/\sqrt{2}\,, 1]$ we use the ring-like equation
(\ref{3.30}), while for $l\in[1\,, \sqrt{3/2}]$ the two-discs
equation (\ref{3.34}) is used. The integrated area
is $1$. $\hfill {\Box\,}$
%
%#####################################################################

\vspace*{0.4cm}
\noindent
{\bf The special cases with $b=0$}

\noindent
In these cases the axis $\L$ of the screw motion contains the centers 
$C$ and $C_g$ of the solid balls $\B$ and $\Bg$; 
the intersections $\BUCr$ and $\BgUCr$ are ring-like and have
constant width, as well as the combined intersections $\BUBgUCr$.  
{}From (\ref{3.2}) we have $m=t$, then (\ref{2.3}) gives 
\bea                                       \label{3.35}
\VgB=\frac{\pi}{12}(2a-t)^2 (4a+t) \hspace{0.7cm} [b=0]
\eea 
provided $t<2a$; this sine qua non is assumed whenever $b=0$. 
The equations (\ref{3.17})-(\ref{3.20}) simplify to  
\bea                                      \label{3.36}
\hspace{-0.5cm}
z_1 = -z_2 \!\!&=&\!\! \sqrt{a^2-r^2}\, , \nonumber \\
z_3 &=& t+z_1\, , z_4 = t-z_1\, ,
\eea 
provided $r<a$; since we assumed $t>0$, we have $min(z_1,z_3)=$
$\sqrt{a^2-r^2}$ and $max(z_2,z_4)=t-\sqrt{a^2-r^2}$, while the
condition $z_1>z_4$ implies
\bea                                              \label{3.37}
r < r_{max}=\sqrt{a^2-t^{2}/4} \, \, . 
\eea
The area of the combined intersection is then 
\bea                                              \label{3.38}
\hspace{-0.3cm}
\SgBr=2\pi r(2\sqrt{a^2-r^2}-t) \;\;\;\; [b=0]
\eea
whenever positive, otherwise it is zero. 
Finally the probability density is, when $b=0$, 
\bea                                              \label{3.39}
\hspace{-1.5cm} && \PgBl=\frac{dr}{dl}\frac{\SgBr}{\VgB}
\nonumber \\
\hspace{-1.5cm} && = \frac{6\,l\,(2\sqrt{a^2-r^2}-t)}
{(2a-t)^{2} (4a+t) } \: {{\csc}^2} \omega/2
\eea
if positive, otherwise it is zero; in (\ref{3.39}) one clearly has to
substitute $r$ for its expression eq.(\ref{3.7}). 
As a matter of fact we have $\PgBl\ne 0$ when $b=0$ only for 
$l_{min}<l<l_{max}$, where 
\bea                                                 \label{3.40}
\hspace{-1.5cm} && l_{min} = t \, , \nonumber \\
\hspace{-1.5cm} && l_{max} = \sqrt{t^2{\cos}^2 \omega/2
+ 4a^2{\sin}^2{\omega/2}} \, ,
\eea
and we have in these extreme limits 
\bea                                               \label{3.41}
\hspace{-1cm} && {\cal P}_g^{\cal B}(l_{min}) =
6t(2a-t)^{-1}(4a+t)^{-1}{\csc}^2{\omega/2} \, , \nonumber \\
\hspace{-1cm} && {\cal P}_g^{\cal B}(l_{max})=0 \, ;
\eea
a sample graph of $\PgBl$ for screw motions with $b=0$ is given in
figure 11. 

\vspace*{1.0cm}
%\hspace*{0.3cm}
\psfig{file=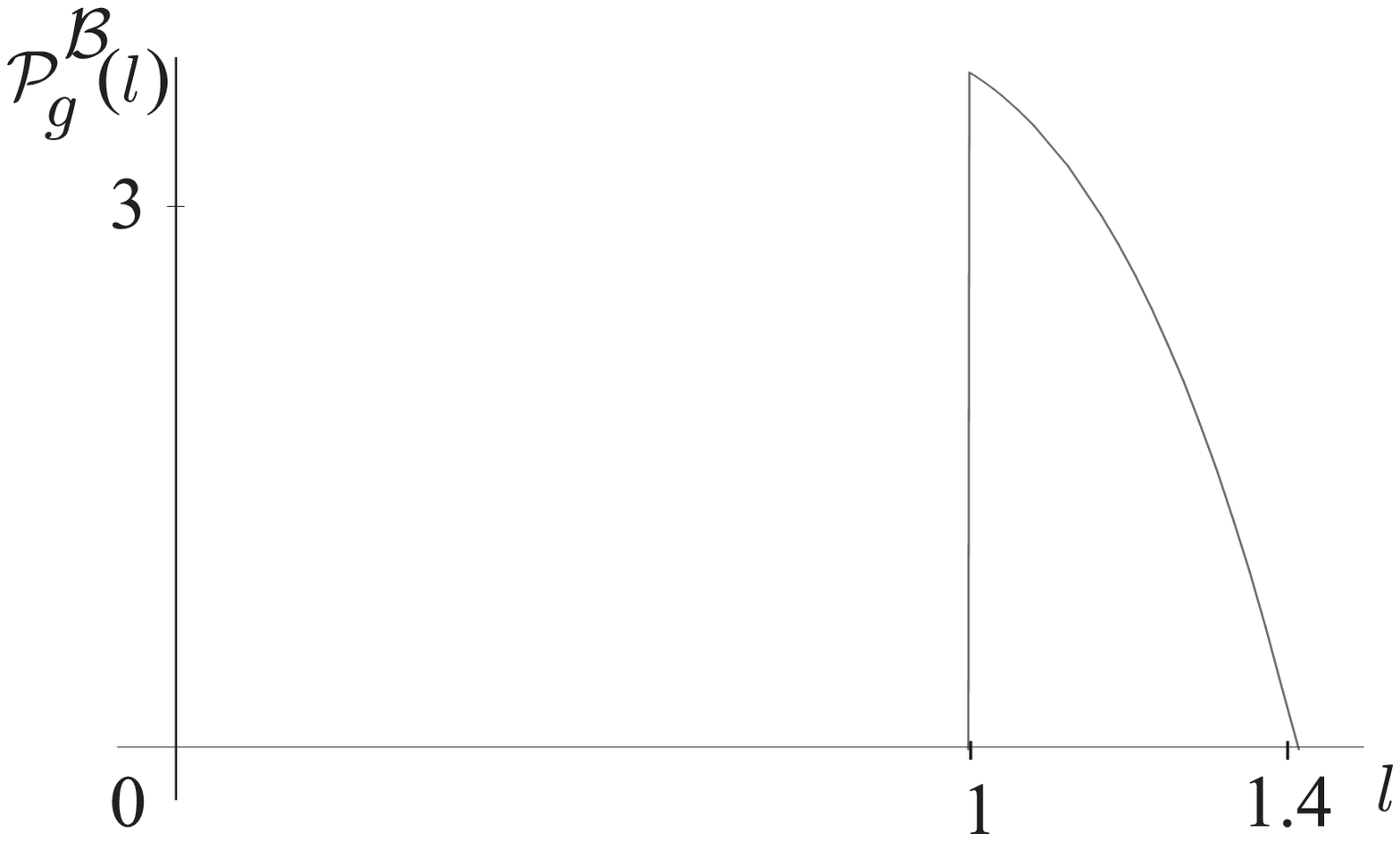,width=7.4cm}   %Figura111111111111111111111111

\vspace*{0.7cm}
\noindent
{\bf Figure 11} The probability density $\PgBl$ for the screw motion
of the solid ball $\B$ with  $b=0,\, a=1/\sqrt{2},\, t=1$\,, and
$\omega=\pi$\,, eq.(\ref{3.39}).
The integrated area is $1$. Note that $l_{min}=1$\,,
$l_{max}=\sqrt{2}$\,, and
${\cal P}_g^{\cal B}(1)=6(3+\sqrt{2})/7\sim3.8$\,. $\hfill {\Box\,}$
%
%#####################################################################

\noindent
{\bf Rotations}

\noindent
A rotation is the limit of a screw motion when the translation tends
to zero, so it is formally described by simply setting $t=0$ in the
appropriate preceding equations.
In particular, eq. (\ref{3.4}) becomes $l=2r\sin\omega/2$.
Rotations of the solid sphere $\B$ are classified into two
categories:

\noindent
(i) if $a<b$ then $\B$ is exempt of fixed points;

\noindent
(ii) if $b<a$ then the axis $\L$ traverses $\B$, which has 
fixed points. 

In category $a<b$ the intersection condition (\ref{3.3}) clearly must
be satisfied (see figure 12), 
\bea                                               \label{3.42}
b\,\sin\omega/2<a \, \, .
\eea
\vspace*{0.6cm}
\hspace*{1.0cm}
\psfig{file=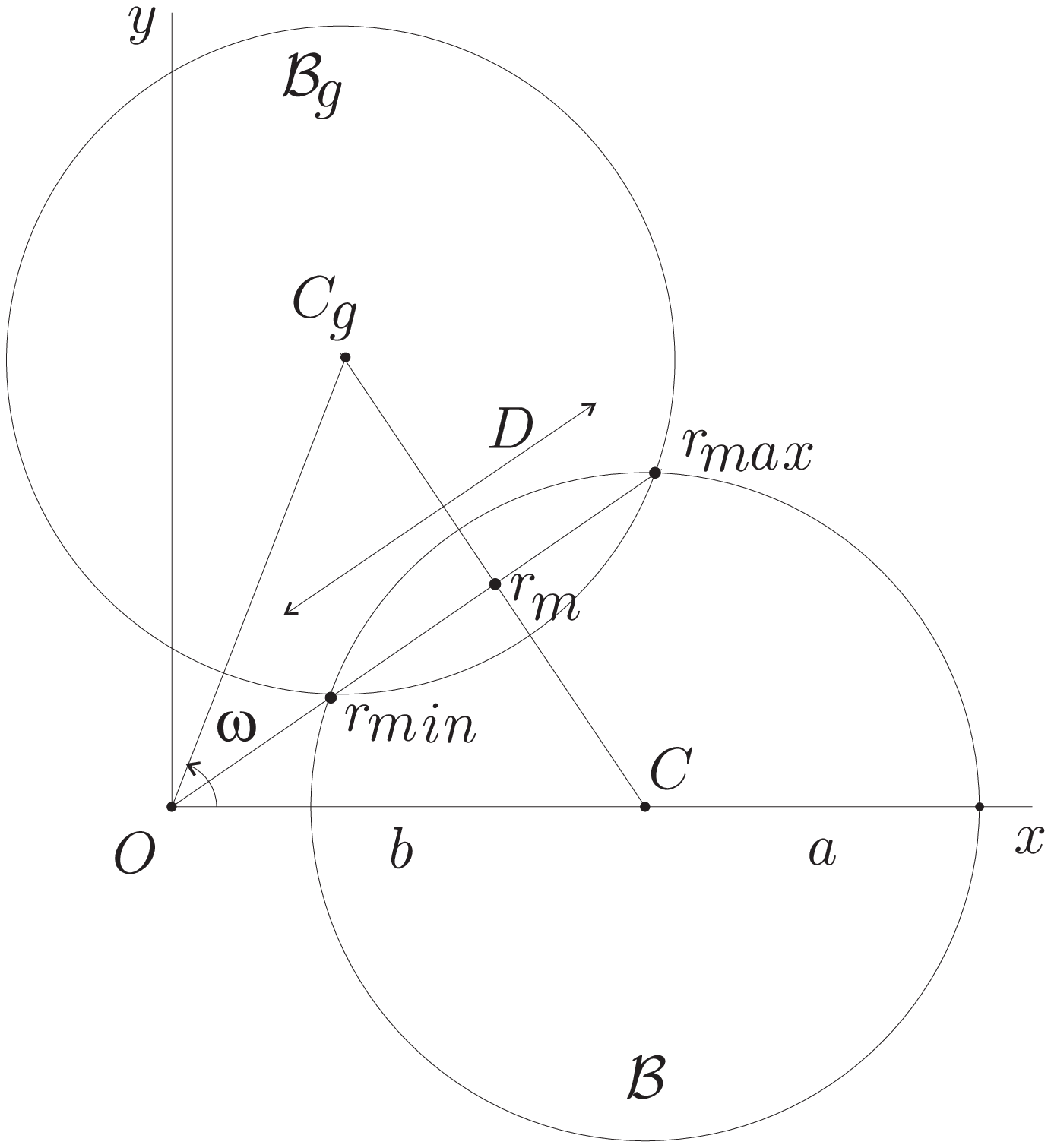,width=5cm}   %Figura1212121212121212121212

\noindent {\bf Figure 12} The solid lens $\BUBg$ when the solid ball $\B$
rotates $\omega$ around the $z$-axis and  $b\sin\omega/2<a<b$\,;
the lens lays between the radial positions $r_{min}$ and $r_{max}$
given by eq. (\ref{3.43}). $\hfill {\Box\,}$

The lens $\BUBg$ lays between the radial positions 
\bea                                             \label{3.43}
&& r_{min}=r_m-D/2 \,, \,\, r_{max}=r_m+D/2 \, , \;\;\;\;
\nonumber \\
&& r_m=b\,\cos\omega/2 \, , 
\eea
where $r_m$ is the radial coordinate of the center of the lens and $D$
is its diameter (\ref{2.3}) with 
\bea                                            \label{3.44}
m=2\,b\,\sin\omega/2 \, \, .
\eea
All intersections $\BUCr$ (and $\BgUCr$) are 
topological discs. Since $z_1>z_4$ and $z_4=-z_3$ in rotations, the
area~(\ref{3.21}) has the simpler expression 
\bea                                            \label{3.45}
\SgBr=2r\int_{-\phi_{max}}^{\phi_{max}} min(z_1,z_3)d\phi  \,\, .
\eea 
When $b\,\sin\omega/2<a<b$ and $r_m-D/2<r<r_m+D/2$ we obtain 
\bea                                              \label{3.46}
\!\!\!\!\! \SgBr \!=\!
8rf\! \Bigl[\!E(1/k,k)-E(\sin\omega/4,k)\!\Bigr] ,
\eea
with $f(r)$ and $k(r)$ given in (\ref{3.23}). The probability density 
(\ref{3.8}) is easily written if we take in succession
$l=2r\sin\omega/2$\,,
then $\QgBr=\SgBr/\VgB$ with (\ref{3.46}), finally $\VgB$ in (\ref{2.3}) 
with $m=2\,b\,\sin\omega/2$\,; in figure 13 a sample graph of
$\PgBl$ for rotation with $a<b$ is given. 

\vspace*{0.4cm}
\hspace*{0.2cm}
\psfig{file=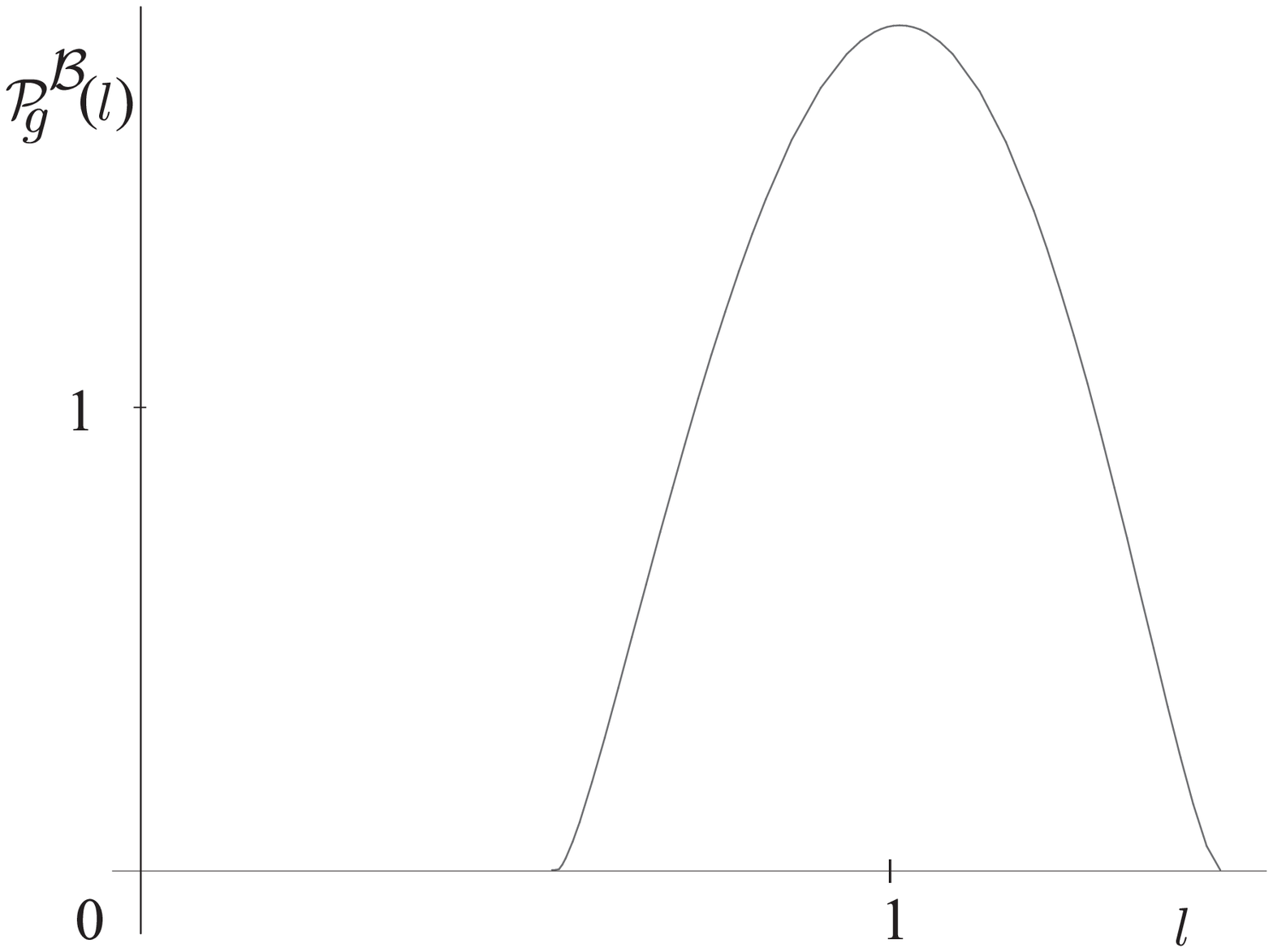,width=7cm}   %Figura13131313131313131313

\vspace*{0.5cm}
\noindent {\bf Figure 13} The probability density $\PgBl$
for a rotation $\omega$ of the solid ball $\B$ with radius $a<b$\,.  
Here $a=1\,, b=2\,,\; $and$\; \omega=\pi/6$\,.
The integrated area is 1. $\hfill {\Box\,}$

In category $b<a$ the intersections $\BUCr$ are either 
ring-like or disc-like, and the combined intersections 
$\BUBgUCr$ can be of three types according to 
the relative values of $a, b, \omega,$ and $r$ (see figure 14): 

\noindent
(i) a ring, when $0<r<a-b$\,; then
\bea                                                \label{3.47}
\hspace{-1.0cm} && \SgBr=8rf\,\Bigl[\,2\,E(k) \nonumber \\
\hspace{-1.0cm}
&& - E(\sin\omega/4,\,k) -E(\cos\omega/4,\,k)\,\Bigr] \, ;
\eea
(ii) a pair of discs, when $a-b<r<D/2-r_m$\,; then 
\bea                                                \label{3.48}
\hspace{-1.0cm} && \SgBr=8rf\,\Bigl[\,2\,E(1/k,\,k) \nonumber \\
\hspace{-1.0cm} && - E(\sin\omega/4,\,k)
-E(\cos\omega/4,\,k)\,\Bigr] \, ;
\eea
(iii) a disc, when $D/2-r_m<r<D/2+r_m$\,; then 
\bea                                                \label{3.49}
\!\!\!\!\! \SgBr\!=\!
8rf\! \Bigl[E(1/k,k)\!-\!E(\sin\omega/4,k)\Bigr] .
\eea
%\vspace*{0.4cm}
\psfig{file=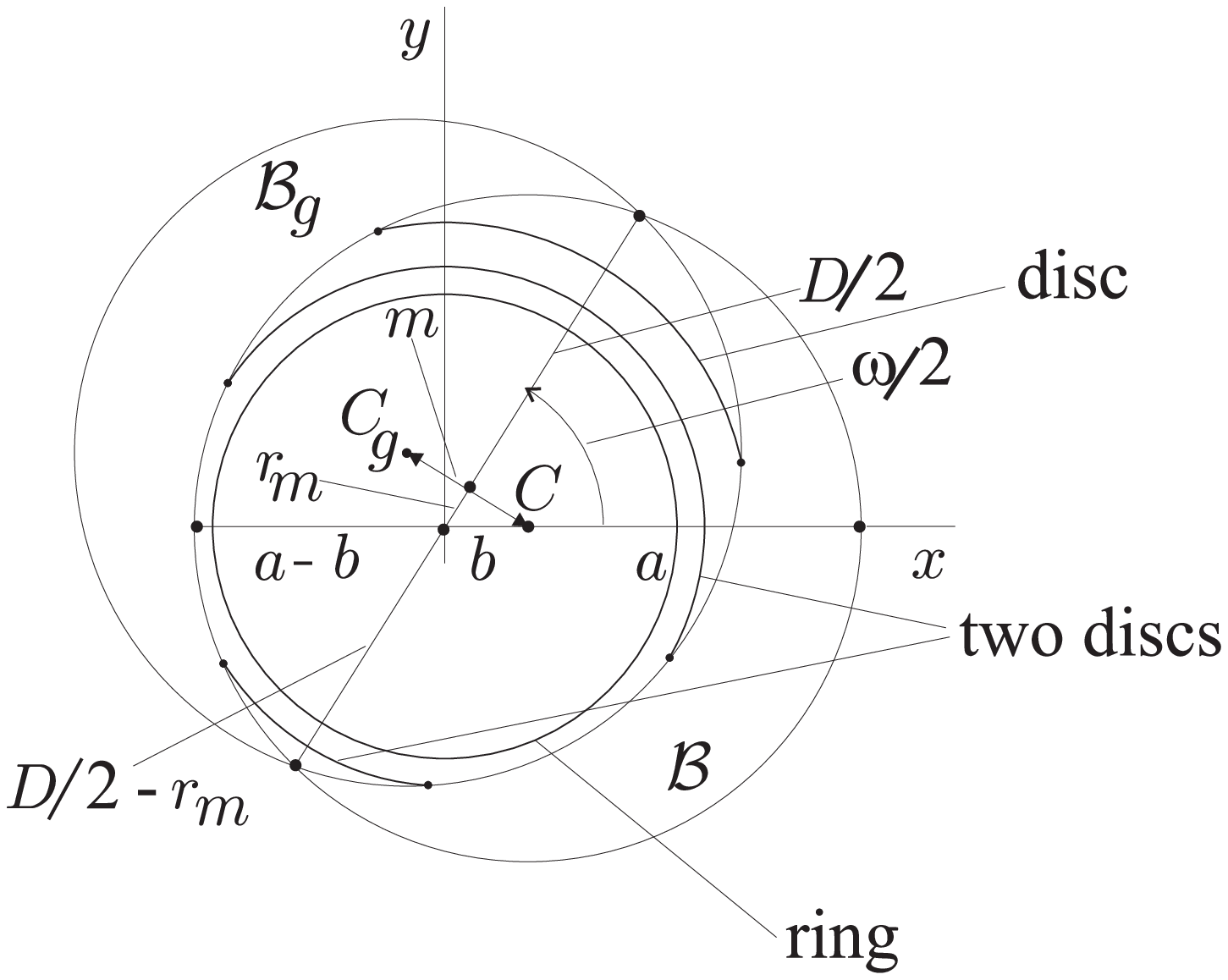,width=7cm}   %Figura14141414141414141414

\vspace*{0.3cm}
\noindent {\bf Figure 14} The solid ball $\B$ with center $C$
and radius $a>b$ rotates $\omega$ around the $z$-axis and gives 
the new ball $\Bg$ with center $C_g$\,. 
The solid lens $\BUBg$ is intersected by the cylindrical surface
$\Cr$ with radius $r$ and axis along the $z$-axis. The intersection
$\BUBgUCr$ is a ring if $r$ is small $(0\leq r<a-b)$, is a pair of
discs for $a-b<r<D/2-r_m$, and is a single disc when $r$ is larger 
$(D/2-r_m<r<D/2+r_m)$. $\hfill {\Box\,}$

In figures 15 and 16 sample graphs of the probability $\PgBl$ 
for rotation with $b<a$ are given, for the entire interval 
$0 \! < \! r \! < \! D/2 \! + \! r_m$\,. 

\vspace*{0.4cm}
\psfig{file=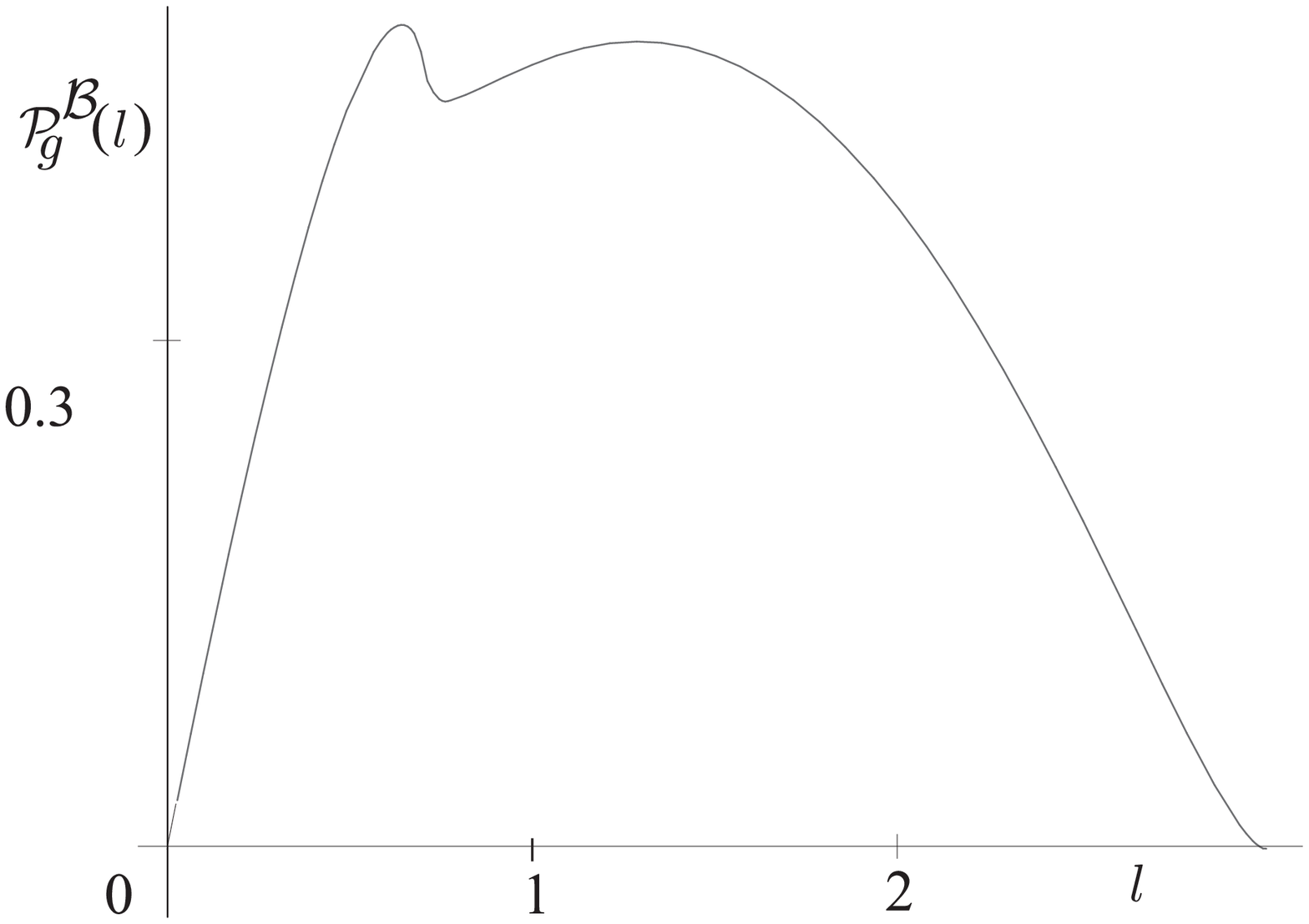,width=7cm}   %Figure 15 15 15 15 15 15 15 15

\noindent {\bf Figure 15} The probability density $\PgBl$ for a 
(pure) rotation of the solid ball $\B$ with radius $a>b$\,.  
Here $a=2\,, b=1.3$\,, and $\omega=\pi/3$.
The integrated area is 1. $\hfill {\Box\,}$

\vspace*{0.4cm}
\hspace{1.0cm}
\psfig{file=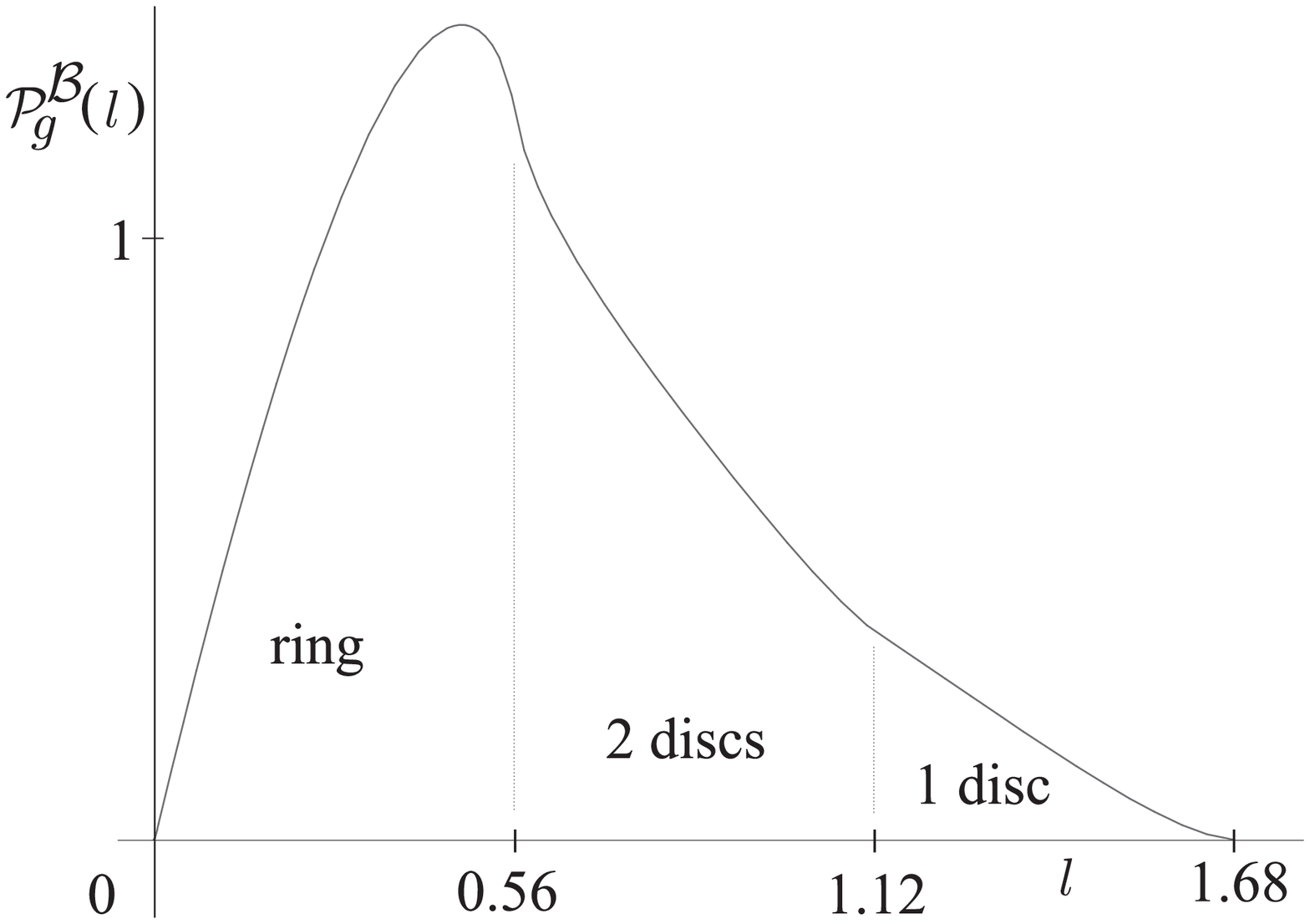,width=7cm}   %Figure 16 16 16 16 16 16 16 16

\noindent {\bf Figure 16} The probability density $\PgBl$ for a 
(pure) rotation of the solid ball $\B$ with radius $a>b$\,.  
Here $a=1\,, b=5/7$\,, and $\omega=2\,{\sec}^{-1}5$. The region where
each of the three expressions (\ref{3.47})-(\ref{3.49}) for $\SgBr$
is used is indicated. The integrated area is 1. $\hfill {\Box\,}$
%
%#####################################################################

\vspace*{0.3cm}
\noindent
{\bf Rotations with $b=0$}

\noindent 
In the category  $b<a$ the special cases $b=0$ deserve a few words. 
The solid spheres $\B$ and $\Bg$ coincide, and we easily 
find that 
\bea                                              \label{3.50}
&& l=2r\sin\omega/2 \, , \,\,\,
\SgBr= 4\pi r\sqrt{a^2-r^2} \, , \nonumber \\
&& \VgB=4\pi a^3/3 \, .
\eea
The probability density is then 
\bea                                             \label{3.51}
\PgBl=\frac{3}{8}
\frac{l\sqrt{4a^2\sin^2\omega/2-l^2}}{a^3\sin^3\omega/2}
\eea
whenever $0\leq l\leq 2a\sin\omega/2$ \,; a sample graph is given in
figure 17.

\vspace*{0.4cm}
\psfig{file=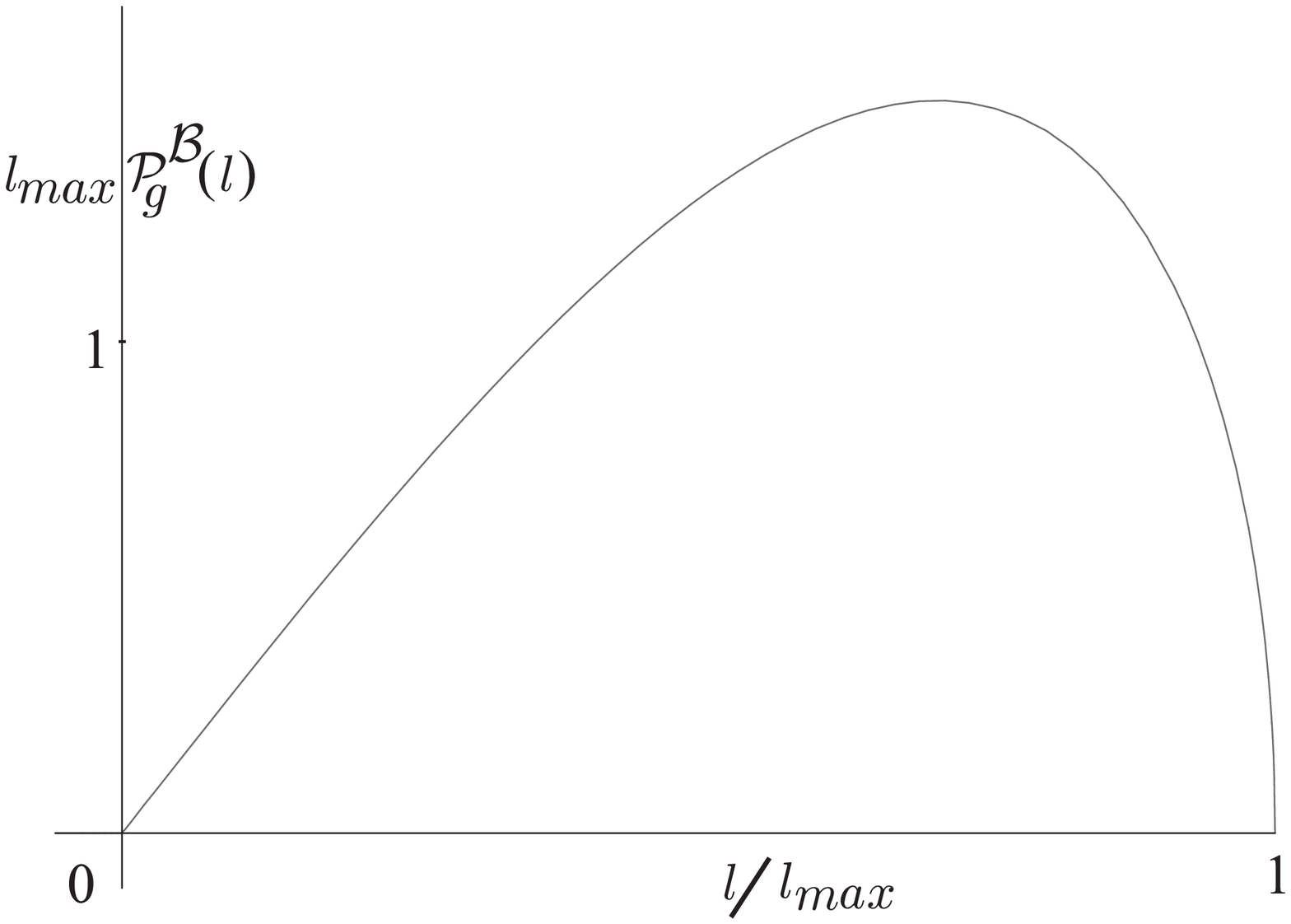,width=8cm}   %Figure 17 17 17 17 17 17 17

\vspace*{0.3cm}
\noindent {\bf Figure 17} The probability density $\PgBl$ for a
(pure) rotation with $b=0$.
Here $a=1$\,, and  $\omega=\pi/6$\,.
The integrated area is 1. $\hfill {\Box\,}$
%
%#####################################################################
\section{Reflections and glide reflections} \label{sec4}
\setcounter{equation}{0} 
In $E^3$, let a solid ball $\B$ with radius $a$ have its center $C$ 
at the cartesian position $(b, 0, 0)$. 
Next consider a glide reflection $g$ on the plane $x=0$ 
(the plane $\X_0$) with nonzero translation $t$ in the direction $+z$. 
The center $C_g$ of the glide reflected ball $\Bg$ is at the cartesian 
position $(-b, 0, t)$, and the separation $m$ between $C$ and $C_g$ is 
\bea                                            \label{4.1}
m=\sqrt{4b^2+t^2}\, .
\eea
The condition of intersection $\BUBg \neq \phi$ implies 
the constraint 
\bea                                            \label{4.2}
4b^2+t^2 < 4a^2
\eea
between the three independent parameters $a, b,$ and $t$; a fortiori 
$b<a$, so both balls intersect the plane $\X_0$; see figure 18. 

\vspace*{0.4cm}
\hspace{1.0cm}
\psfig{file=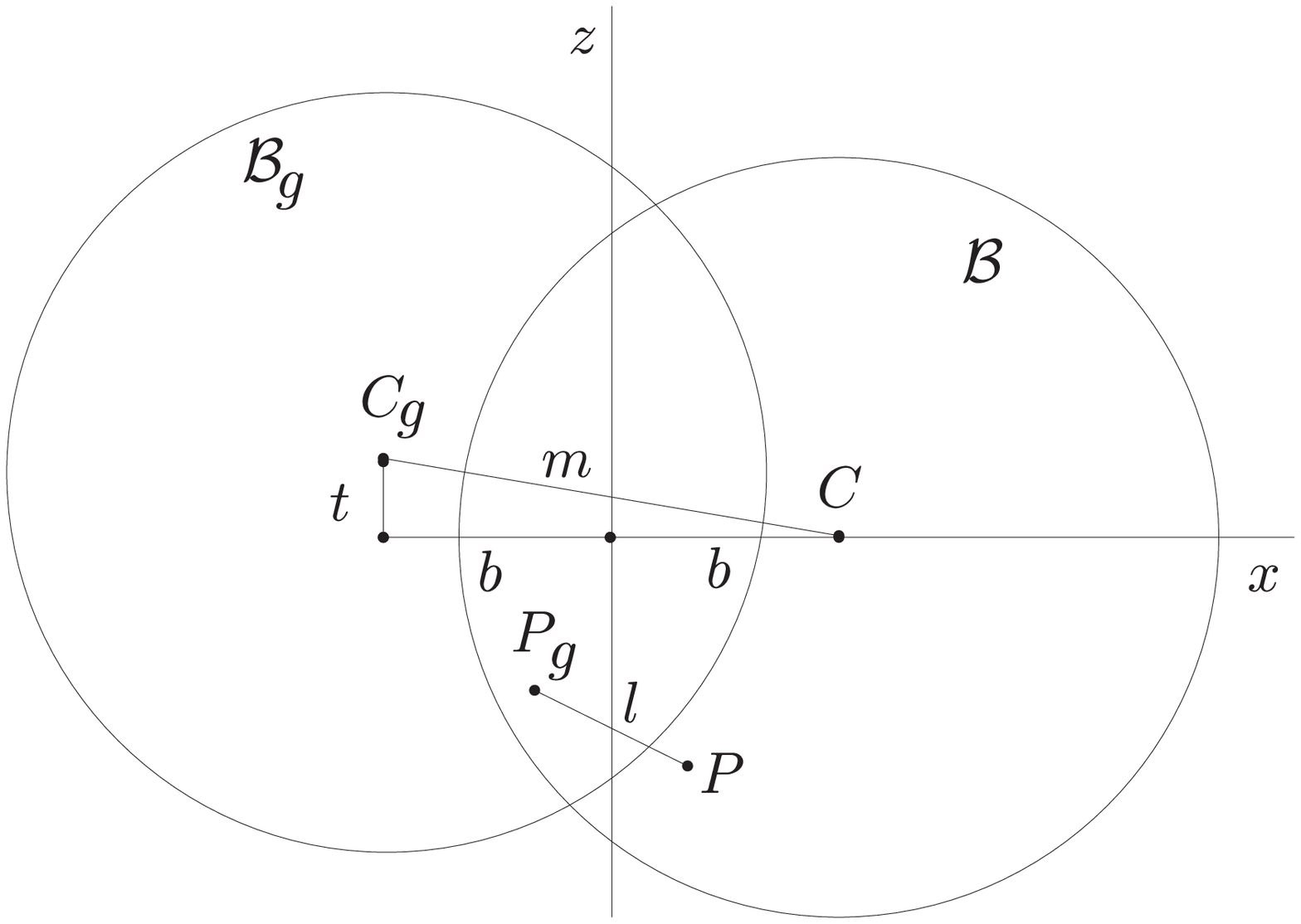,width=7cm}   %Figure 18 18 18 18 18 18 18

\noindent {\bf Figure 18} The solid ball $\B$ with center $C$ 
and radius $a$ is first reflected on the plane $x=0$ a distance 
$b<a$ apart, then translated $t$ upwards. $\hfill {\Box\,}$

Now randomly choose a point $P=(x, y, z)$ of $\B$, such that 
$P_g=(-x, y, z+t)$ is also in $\B$; clearly 
$P\in\BUBg$. $l$ being the separation from $P$ to $P_g$, 
we ask for the probability density $\PgBl$ as described in sec. 2; 
we readily find 
\bea                                            \label{4.3}
\PgBl &=& 2 \, \frac{area(\BUBgUXx)}{vol(\BUBg)} \nonumber \\
      &=& 2 \, \frac{\SgBx}{\VgB} \, ,
\eea
where $\X_x$ is one of the two planes 
\bea                                            \label{4.4}
x=\pm\frac{1}{2}\sqrt{l^2-t^2} \, ;
\eea
the multiplying factor 2 in (\ref{4.3}) accounts for these two
possibilities, and the volume $\VgB$ is given in (\ref{2.3})\,.

The intersections $\BUBgUXx$ are of two types, depending on whether
or not the plane $\X_x$ intersects the equator ${\cal E}$ of the lens 
$\BUBg$ (see figure 19):

\vspace*{0.4cm}
\psfig{file=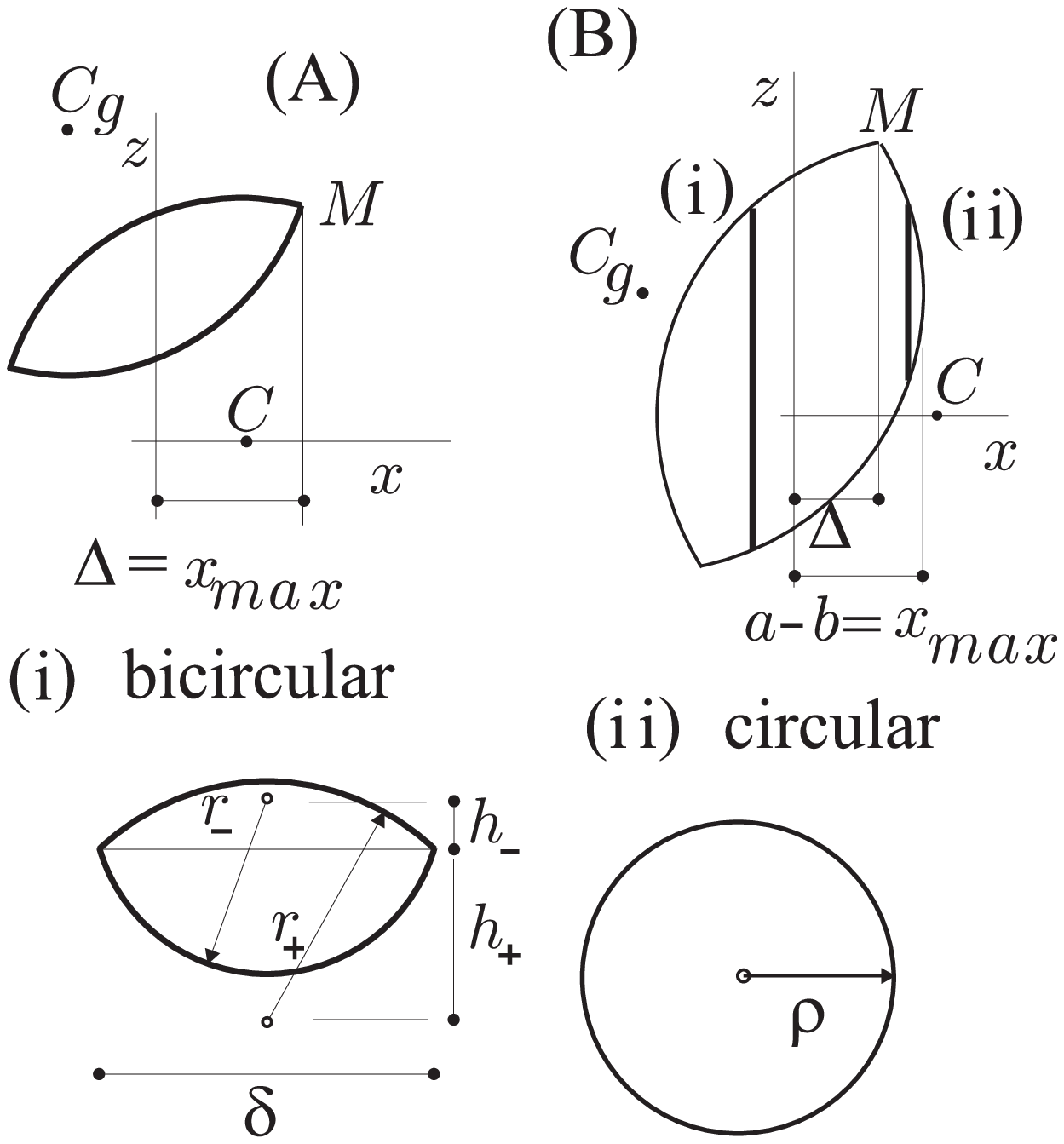,width=7cm}   %Figure 19 19 19 19 19 19

\vspace*{0.5cm}
\noindent
{\bf Figure 19} The solid lens $\BUBg$ and the two types of
intersections $\BUBgUXx$ when the isometry is a glide reflection. 
In the equator ${\cal E}$ of the lens, $M$ is the point placed
farthest from the reflector plane  $x=0$\,. 
In ({\bf A}), $M$ is below the center $C_g$ of the glide reflected
ball $\Bg$\,, then $x_{max}=x_M=\Delta$ given in (\ref{4.8}); only
bicircular sections (i) are possible. 
In ({\bf B}), $M$ is above $C_g$\,, then $x_{max} = a-b > x_M$
and two types of vertical cross-sections 
can occur: \,\, (i) bicircular, and (ii)
circular. $\hfill {\Box\,}$

\vspace*{0.3cm}
\noindent
(i) a bicircular disc, when $\X_x\cap{\cal E}\neq\phi$\,; the disc 
is enclosed by two circles with unequal radii 
\bea                                            \label{4.5}
r_{\pm}(x) = \sqrt{a^2 - (b \mp x)^2}\, .
\eea
It has area
\bea                                           \label{4.6}
\SgBx \!=\! r_{-}^2 \cos^{-1} \!\frac{h_{-}}{r_{-}}
+ r_{+}^2\cos^{-1} \!\frac{h_{+}}{r_{+}}
- \frac{t \delta}{2} ,
\eea
whenever $|x|<\Delta$\,, where 
\bea                                                  \label{4.7}
h_{\pm}(x) = t/2 \pm 2\,b\,x / \,t \,,
\eea
\bea                                                  \label{4.8}
\hspace{-1cm} &&\Delta
= \frac{t}{2}\,\sqrt{\frac{a^2}{b^2+t^2/4}-1}\,,
\nonumber \\
\hspace{-1cm} &&\delta(x) = 2\,\sqrt{(1+4b^2/t^2)(\Delta^2-x^2)}\,.
\eea
(ii) a circular disc with radius 
\vspace{-0.3cm}
\bea                                                  \label{4.9}
\rho=\sqrt{a^2+(b+|x|)^2}\,,
\eea

\vspace{-0.3cm}
\noindent
which appears only when
\vspace{-0.3cm}
\bea                                                  \label{4.10}
\!\!\!\!\!q \!=\! 4b(a\!-\!b)-t^2 \!>\! 0, \,\,
\Delta \!<\! |x|<a-b \,;
\eea

\vspace{-0.3cm}
\noindent
the area of the disc is clearly $\pi\rho^2$\,.

Collecting together preceding terms we finally obtain 
\vspace*{-0.3cm}
\bea                                                  \label{4.11}
\hspace{-1.5cm}
&& \PgBl=\frac{l}{\sqrt{l^2-t^2}}
\frac{1}{\VgB}\,\Bigl[\Theta(\Delta-|x|)\,\SgBx \nonumber \\
\hspace{-1.5cm}
&& + \Theta(q)\Theta(|x|-\Delta)\Theta(a-b-|x|)\pi\rho^2\,\Bigr]\,,
\eea

\vspace{-0.3cm}
\noindent
valid for $t < l\leq l_{max}$ where
\vspace{-0.3cm}
\bea                                                  \label{4.12}
\hspace{-1cm}
&& l_{max}=\sqrt{t^2+4(a-b)^2}\,\Theta(q) \nonumber \\
\hspace{-1cm}
&& +\frac{at}{\sqrt{b^2+t^2/4}}\,\Theta(-q)\,.
\eea

\vspace{-0.3cm}
\noindent
A graph of $\PgBl$ is given in figure 20.

\vspace*{0.3cm}
\psfig{file=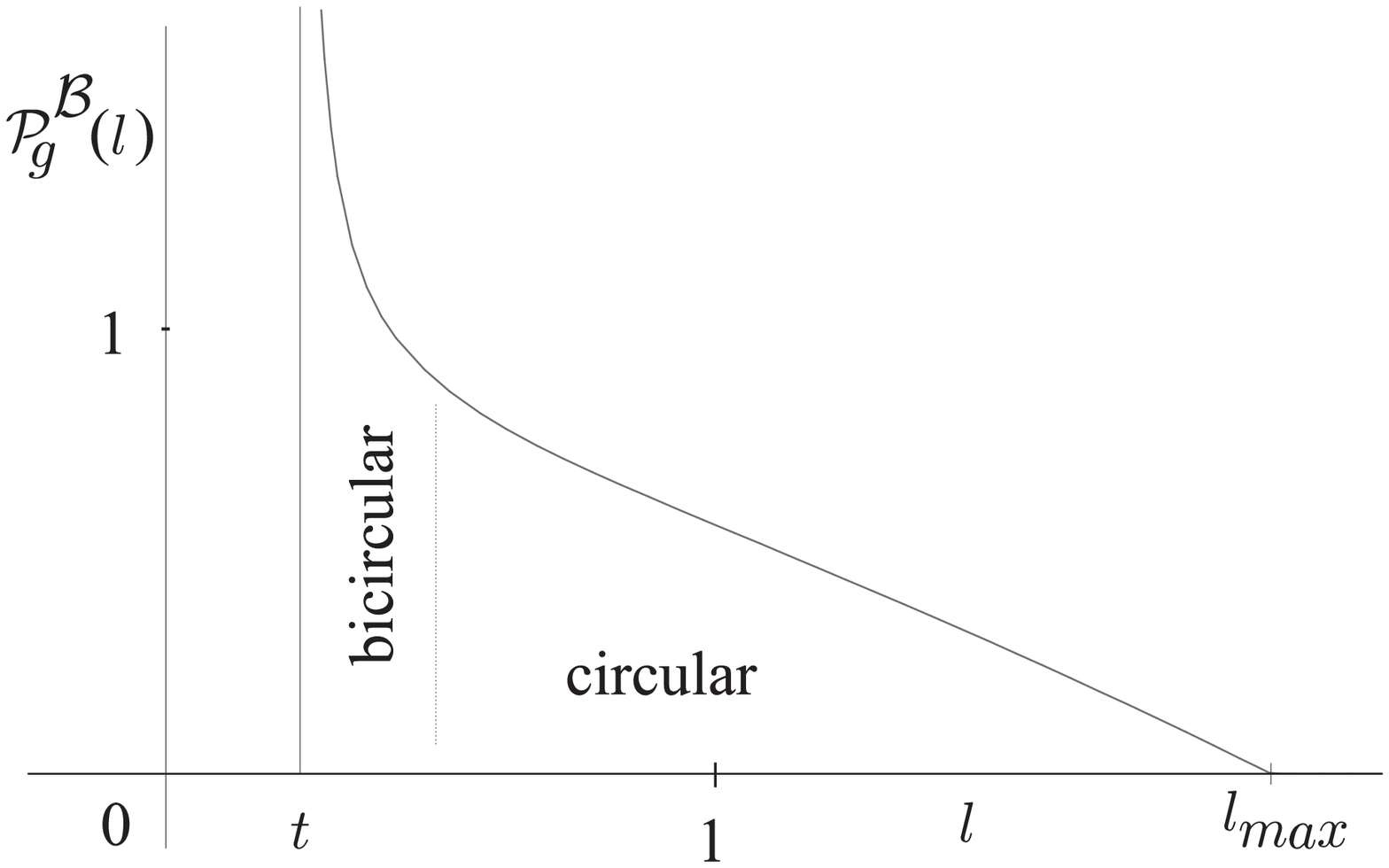,width=7cm}   %Figure 20 20 20 20 20 20 20

\vspace*{0.5cm}
\noindent
{\bf Figure 20} The probability density $\PgBl$ for a solid ball
under a glide reflection, eq.(\ref{4.11}). Here $a=2\,, b=1$\,, and
$t=0.25$\,.
The regions where the sections $\BUBgUXx$ are bicircular or circular
are displayed.
The function diverges when $l\rightarrow t$\,, nevertheless the
integrated area from $t$ to $l_{max}$ is
finite. $\hfill {\Box\,}$
%
%#####################################################################

\vspace*{0.3cm}
\noindent
{\bf Reflections}

\noindent
(Pure) reflections are glide reflections whose
translation is $t=0$\,.

\psfig{file=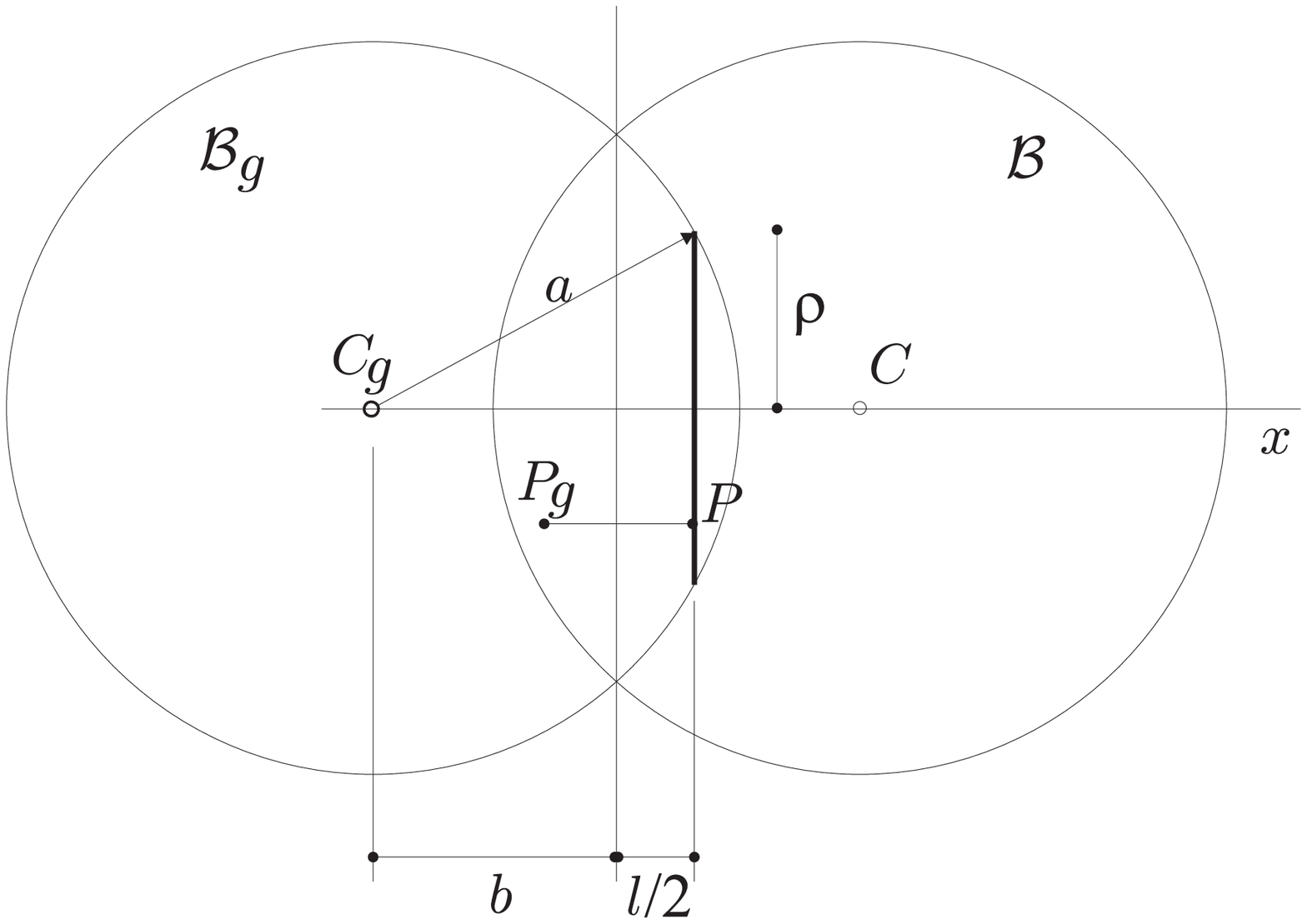,width=7cm}   %Figure 21 21 21 21 21 21 21

\noindent {\bf Figure 21} The solid lens $\BUBg$ when the isometry 
$g$ is a (pure) reflection on the plane $x=0$\,. The sections 
$\BUBgUXx$ are circles with variable radius
$\rho$\,. $\hfill {\Box\,}$

\vspace*{0.3cm}
As is evident from figure 21,
the probability density $\PgBl$ is proportional to the area of a
disc with radius
\bea                                                \label{4.13}
\rho=\sqrt{a^2-(b+l/2)^2}\,,
\eea
and is given by 
\bea                                                \label{4.14}
\PgBl
= \frac{3}{2}\frac{a^2-(b+l/2)^2}{(a-b)^2(2a+b)} \, ,
\eea
for $0 \leq l \leq 2(a-b)$.
A graph is presented in figure 22.

\vspace*{0.4cm}
\hspace*{1.0cm}
\psfig{file=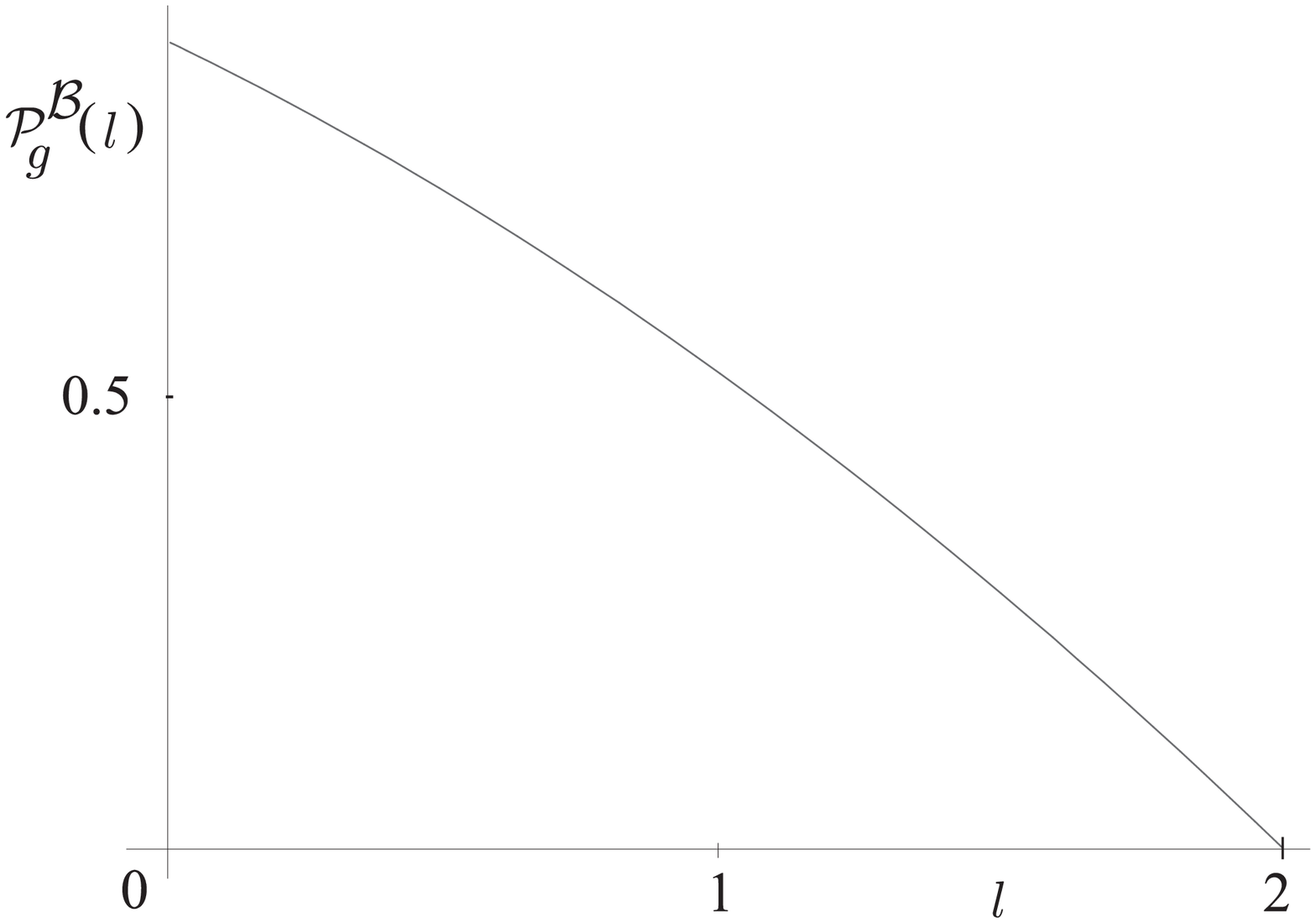,width=7cm}   %Figure 22 22 22 22 22 22 22 

\noindent {\bf Figure 22} The probability density $\PgBl$ for a solid 
ball when the isometry $g$ is a (pure) reflection, eq.(\ref{4.14}). 
Here $a=2$\, and $b=1$\,.
The integrated area is 1. $\hfill {\Box\,}$
%
%
%#####################################################################
\section{Discussions} \label{sec4}
\setcounter{equation}{0} 
\vspace*{-0.2cm}
We initially aimed to write out one single expression for the
probability density $\PgBl$ for screw motions of solid balls, valid
for whatever values of the {\it four} parameters $a, b, t$\,, and
$\omega$.
However, we soon found that such expression would demand a quite large 
number of step functions to account for all sort of possibilities. 
Since in practice the isometries are dealt with one at each time, 
we found more appropriate to present a simple method to have the
exact $\PgBl$ for each individual screw motion with {\it fixed}
values of the four parameters.
Nevertheless, for those isometric motions of solid balls described by 
{\it three} or less free parameters the exact expression for $\PgBl$ 
for {\em any} euclidean isometry is short enough and was displayed. 

As promised in the Introduction, we exhibited the analytic
counterpart $\PgBl$
of the computer simulations of pair separations histograms of the
euclidean
isometries in cosmic crystallography thus far obtained in the
literature.
The graph of $\PgBl$ in figure 10 corresponds to the isometries $b$
and $c$
in the Fagundes and Gausmann~\cite{CESU} study, or equivalently the
isometries $\beta$ and $\delta$ in Gomero~\cite{FPGD}. 
The discontinuity in $l\sim{0.7}$ is not observed in the two upper
figures 1 (Universe E4) of~\cite{CESU} due to the strong statistical
noise present in these histograms; nevertheless it is clearly seen
in the position $s=l^2=0.5$ in the {\em mean} histogram 5b
of~\cite{spikesII} as well as in the position $l\sim{0.7}$ in the
mean histograms 1a, 1b, 2b, 4a, and 5 of \cite{tsct}.
Similarly, figure 11 corresponds to the isometry $a$ of \cite{CESU}
and $\alpha$ in~\cite{FPGD} and \cite{spikesII}; the discontinuity
in $l=1$ has their counterparts again in the histogram 5b
of~\cite{spikesII} and in the histograms 1a, 1b, 2a, 4a, and 5
of~\cite{tsct}.

In contrast with the screw motions (figures 10 and 11), the pure
rotations (figures 13, 15, and 16, all with $t=0$) do not show
discontinuity of $\PgBl$\,. 
Oppositely to figure 13, where $a<b$\,, figures 15 and 16
correspond to $a>b$\,, so the ball $\B$ now has fixed points and the
graph of $\PgBl$ effectively starts from $l=0$. 
The strange-looking graph in figure 15 was confirmed in a computer 
simulation; the irregular behavior near $l=0.7$ corresponds to the
narrow $r$-interval where the pair-of-discs combined intersection
eq.(\ref{3.48}) occurs.

In figure 16 we have chosen values for $a, b$, and $\omega$ such
that the three types of combined intersection (ring, pair of discs,
and one disc) have equal range in the $l$ scale. 
{\it En passant}, the pair-of-discs $l-$range is now wide, and does
not originate a bump as did in figure 15. 

Figure 17 corresponds to rotation of the solid ball $\B$ around a
diameter ${\cal D}$; from (\ref{3.51}) we find that defining 
$l_{max}=2a\sin\omega/2$ then the graph of $l_{max}\PgBl$ against
$l/l_{max}$ does not depend on $l_{max}$. 
The points of $\B$ along the diameter ${\cal D}$ are fixed under
the isometry, so the graph again effectively starts from the origin.  

Figure 20 corresponds to a glide reflection whose sections $x=const$
in the intersection $\BUBg$ are either bicircular discs (for small
$|x|$) or circular (for larger $|x|$). The minimum displacement $l$
occurs for the points of $\B$ in the intersection with the reflector
plane $x=0$, giving $l_{min}=t$, the translation. 
Since for all points near the reflector plane we have
$l\sim t(1+2|x|^2)$, then these points are displaced almost the same
value $l\sim t$;
as a consequence, $\PgBl$ diverges in the vicinity of $l=t$.
Nevertheless the integrated area is finite, with value 1. 
  
Between $l=t=0.25$ and $l\sim 0.5$ we have bicircular sections 
(\ref{4.6}), while for $0.5<l<l_{max}\sim 2.0$ the sections
$\BUBgUXx$ are circles. 

The transition from a glide reflection to a (pure) reflection is
worth describing: if in figure 20 we continuously displace the
vertical line $l=t=0.25$ towards $l=0$, then the region of
divergence of $\PgBl$ shrinks continuously and disappears when
$l=0$, eventually giving the graph of figure 22.


\begin{thebibliography}{5}
% 
\bibitem{lelalu} R. Lehoucq, M. Lachi\`{e}ze-Rey, and J.-P. Luminet, 
 {\it Astron. Astroph.} {\bf 313}, 339-346 (1996). 
%
\bibitem{spikesI} G.I. Gomero, A.F.F. Teixeira, M.J. Rebou\c{c}as
and A. Bernui, {\it Spikes in cosmic crystallography}, 
gr-qc/9811038 (1998). 
%
\bibitem{spikesII} G.I. Gomero, M.J. Rebou\c{c}as and A.F.F. Teixeira, 
{\it Spikes in cosmic crystallography II: topological signature of compact 
flat universes},  
gr-qc/9909078 (1999). 
%
\bibitem{cc3mpf} A. Bernui and A.F.F. Teixeira, 
{\it Cosmic crystallography: three multipurpose functions},
astro-ph/9904180 (1999).
%
\bibitem{tsct} G.I. Gomero, M.J. Rebou\c{c}as and A.F.F. Teixeira, 
{\it A topological signature in cosmic topology}, 
gr-qc/9911049 (1999). 
%
\bibitem{CESU} H.V. Fagundes and E. Gausmann, 
{\it Phys. Letters A} {\bf 238}, 235-238 (1998). 
%
\bibitem{FPGD} G. Gomero, 
{\it Fundamental polyhedron and glueing data for the sixth
euclidean compact orientable 3-manifold},  
preprint CBPF-NF-049/97 (1997). 
%
\end{thebibliography}
\end{document}